\documentclass[twocolumn]{aastex63}

\usepackage{graphicx}
\usepackage{txfonts}
\usepackage[utf8]{inputenc}
\usepackage{natbib}
\usepackage{float}
\usepackage{mathrsfs}
\usepackage{hyperref}

\newcommand{\rsun}{$R_{\odot}$}

\newcommand{\alfven}{Alfv\'en}

\received{December 7, 2022}
\revised{January 15, 2023}
\accepted{January 24, 2023}
\submitjournal{The Astrophysical Journal}

\shorttitle{Coronal magnetic fields from the 21 August 2017 total solar eclipse}
\shortauthors{Bemporad}

\begin{document}

\title{Coronal Magnetic Fields derived with Images acquired\\
during the 21 August 2017 Total Solar Eclipse}

\author[0000-0001-5796-5653]{A. Bemporad}
\affiliation{Istituto Nazionale di Astrofisica (INAF), Osservatorio Astrofisico di Torino, via Osservatorio 20, 10025 Pino Torinese, Torino, Italy}
\affiliation{Key Laboratory of Dark Matter and Space Astronomy, Purple Mountain Observatory, \\ Chinese Academy of Sciences, 210023 Nanjing, People's Republic of China}
\email{alessandro.bemporad@inaf.it}

\begin{abstract}
The coronal magnetic field, despite its overwhelming importance to the physics and dynamics of the corona, has only rarely been measured. Here, the electron density maps derived from images acquired during the total solar eclipse of August 21st, 2017 are employed to demonstrate a new technique to measure the coronal magnetic fields. The strength of the coronal magnetic fields is derived with a semiempirical formula relating the plasma magnetic energy density to the gravitational potential energy. The resulting values are compared with those provided by more advanced coronal field reconstruction methods based on MHD simulations of the whole corona starting from photospheric field measurements, finding a very good agreement. Other parameters such as the plasma-$\beta$ and \alfven\ velocity are also derived and compared with those of MHD simulations. Moreover, the plane-of-sky (POS) orientation of the coronal magnetic fields is derived from the observed inclination of the coronal features in the filtered images, also finding a close agreement with magnetic field reconstructions. Hence, this work demonstrates for the first time that the 2D distribution of coronal electron densities measured during total solar eclipses is sufficient to provide the coronal magnetic field strengths and inclinations projected on the POS. These are among the main missing pieces of information that limited so far our understanding of physical phenomena going on in the solar corona.
\end{abstract}


\section{Introduction}

The limited knowledge of the coronal magnetic fields is one of the major open problems preventing our full understanding of coronal stationary and dynamic phenomena \citep[see e.g. review by][]{Wiegelmann2017}. For this reason, for decades many different methods have been developed to measure what is "one of the, if not \textit{the} most important physical quantity in the observable solar atmosphere" \citep{Beckers1971}. After many years, the measurement of the coronal magnetic fields became "something of a ''dark energy'' problem for us" \citep{Lin2004}, with the big difference that the dark energy in galaxies and galaxy clusters can be measured, even if we do not have a clear physical explanation for it, while for the magnetic field we know exactly what is it, but we cannot measure it in the solar atmosphere.

As is well known, the magnetic fields are routinely measured in the solar photosphere, thanks to the so-called Zeeman effect \citep[see review by][]{Stenflo1978}, but the application of the same technique for the measurement of coronal fields is very difficult, mostly because of the weakness of these fields (with respect to the photospheric ones), and also because of the thermal line broadening preventing the capability to separate the splitted spectral line components in the corona. The use of the Zeeman effect to measure the weak coronal magnetic fields has been proposed and successfully tested \citep{Lin2000} by exploiting the different circular polarization of the splitted components with spectro-polarimetric measurements \citep{Lin2004}. Many results have shown that these measurements are feasible \citep{Tomczyk2008, Yang2020}, but given the weakness of the circularly polarized coronal emission, they require long exposure times and strong coronal magnetic fields, or exceptional observational conditions \citep[e.g.][]{Kuridze2019}.

Similar measurements have been proposed to estimate the coronal magnetic fields by exploiting another effect modifying the polarized emission of some spectral lines, namely the Hanle effect, but again this technique requires long exposure times and strong magnetic fields \citep[see review by][]{Trujillo2022}. Recently, it was also shown that spectro-polarimetric observations of coronal emission lines can be combined with tomographic reconstructions to get information on the 3D magnetic field configurations \citep[see e.g.][and references therein]{Kramar2016} requiring at least half of solar rotation observations.

Given the complexities in the data of the above analyses, none of the above techniques is currently able to provide routine daily measurements of the coronal magnetic field strength and inclination. In the past, many other techniques have been employed more sporadically, under specific conditions, and provided in general more local estimates. Coronal magnetic field strengths have been estimated for instance from the propagation of EUV waves associated with solar flares, from the dynamic of CME-driven shocks, from type-II radio bursts, from EUV loop oscillations, and also with the coronal seismology technique \citep[see][and references therein]{Gibson2017}.

From the theoretical point of view, at least two approaches are typically used to reconstruct the magnetic fields in the whole corona \citep[see e.g.][]{Riley2006, Mackay2012}. A first approach is based on the assumption of photospheric magnetic fields measured over one full solar rotation, to be used as a boundary condition for magnetic field extrapolations in the solar atmosphere. One of the first methods ever developed is the Potential Field approximation, whose solutions were first provided by \citet{Altschuler1969}, a method still used today to reconstruct the large-scale coronal magnetic field configuration. Further improvements of this method are the linear force-free \citep[e.g.][]{Nakagawa1972} and non-linear force-free extrapolations \citep[e.g.][]{Wu1990}. These methods have been compared to each other by many authors, and the extrapolations have also been compared with the orientations of coronal loops reconstructed in 3D \citep[][]{Wiegelmann2005}.

After many works focusing in the past on magneto-static reconstructions \citep[][]{Low1975, Hundhausen1994}, the second theoretical approach applied today to reconstruct the magnetic fields in the whole corona is based on numerical models starting from the field measurements provided in the photosphere and applying Magneto Hydro-Dynamic (MHD) equations \citep[e.g.][]{Lionello2009, Feng2012}. These models require also the knowledge at a given time of the global photospheric magnetic field, and because (before the launch of Solar Orbiter) these fields were measured only along the Sun-Earth line, hence in the visible solar hemisphere, this also require the development of complex procedures to simulate and reconstruct the evolution of the photospheric fields in the backward solar hemisphere during the solar rotation \citep[see review by][]{Mackay2012}. These MHD models are used for instance to predict, in the occasion of Total Solar Eclipses (hereafter TSEs), the appearance of the corona \citep{Mikic2007}, finding good qualitative agreement between the predictions and the real observations. Based on these global models, or on specific models of magnetic cavities and spheromaks, some authors focused on the forward modeling of the expected polarized emission in specific emission lines \citep{Rachmeler2013, Gibson2016}, to be compared with the real observations looking for similarities or differences as signatures of agreement or disagreement between the reconstructed and the real coronal fields.
\begin{figure*}[t!]
	\centering
	\includegraphics[width=0.8\textwidth]{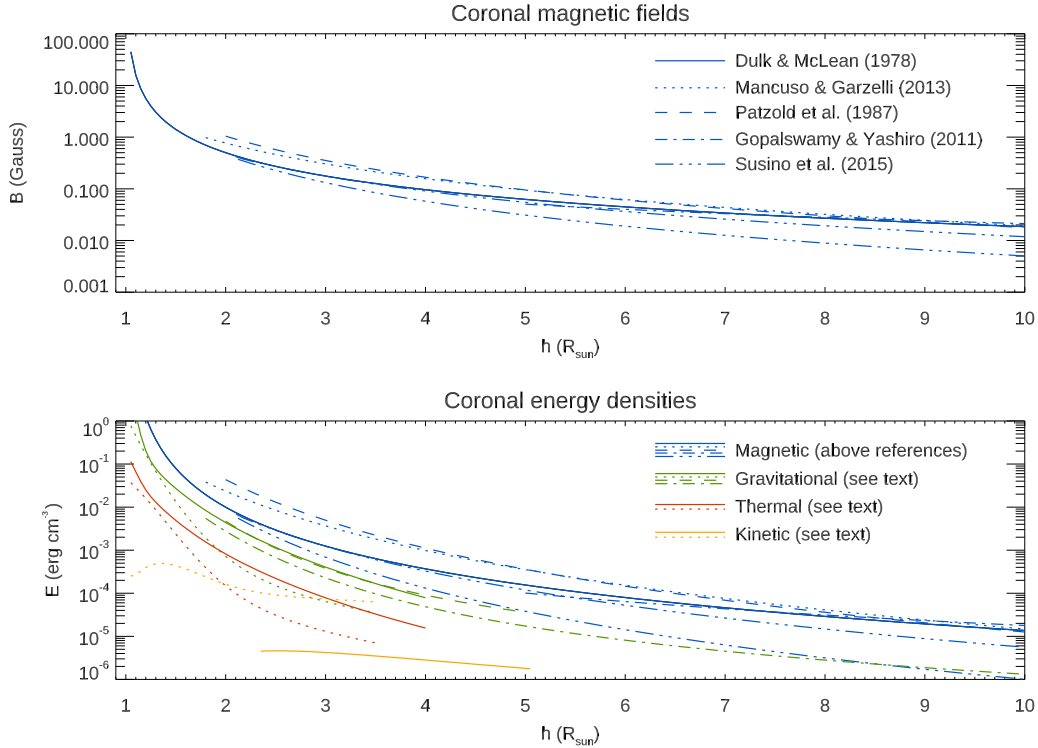}
	\caption{
		Top: a compilation of radial profiles for the coronal magnetic field intensities as measured by different authors with different techniques. Bottom: a comparison between the strength of different coronal energy densities by assuming the above radial profiles for the magnetic field strength (same line styles), and by assuming different radial profiles in the literature (see text) for coronal plasma densities, temperatures, and bulk outflow speeds.  
	}
	\label{Fig1}
\end{figure*}

All the above very complex observational and modeling techniques have been developed over the last decades for a fundamental reason: an existing simple way to routinely measure the coronal magnetic fields has never been identified so far. From the observational point of view, the above mentioned methods (and in particular those aimed at measuring the observed modifications induced by the magnetic field in the polarization of specific coronal emission lines) already provided significant results, and potentially could be applied routinely, but the data inversion is really complex, it requires many assumptions (unknown elemental abundances, validity of ionization equilibrium, line-of-sight  - LOS - integration effects), and also long exposure times to measure the weak polarized emission. On the other hand, from the theoretical point of view, the above mentioned methods to reconstruct the coronal fields reached an unprecedented level of details in the reconstruction of coronal features, but are based on global photospheric fields that are not measured at one single time over the whole Sun, and are based on empirical coronal heating models that are optimized to provide the best match with observations of stationary corona but not able to follow its dynamic.

As I show in this work, a very simple method to measure the coronal magnetic field strengths based on the already available data was missed and never identified so far. Such a method is described here, starting from the results obtained in a previous work \citep{Bemporad2020} and based on the images acquired with cost-effective equipment during the August 21st 2017 TSE. The method shown here is related to a fundamental property of the coronal plasma that has been missed so far, and that I try to explain here based on empirical and theoretical reasoning. Because the method described here is simply based on the analysis of standard coronagraphic images acquired in the visible light, it could be easily applied to convert into a "coronal magnetometer" every already existing or future coronagraph observing the inner corona, such as for instance the ground-based Mauna Loa coronagraphs \citep[][]{Tomczyk2022}, and the space-based ASPIICS coronagraph on-board PROBA-3 \citep[][]{Lamy2017}. In this paper, after general introductory considerations on the distributions of different energies in the solar corona (\S\ref{sec: energies}), I describe the input observational data and the theoretical model employed for the comparison between the measured and the simulated magnetic fields (\S\ref{sec: input}). Then, I introduce the new method for the coronal magnetic field strength measurement (\S\ref{sec: field strength}), and discuss possible theoretical justifications for this method (\S\ref{sec: theoretical}). Finally, the measurement of the coronal field projected inclination is also described (\S\ref{sec: field direction}). Then, the results are discussed and the conclusions are given (\S\ref{sec: summary}).

\section{Coronal plasma energies} \label{sec: energies}

\subsection{Different energy distributions}

The solar coronal plasma contains at different altitudes above the solar surface many different kinds of energies. Some of these energies are better constrained by the observations (for instance the gravitational and thermal energies), while some other energies are more uncertain (for instance the magnetic energy, the wave energy, the bulk kinetic energy, and other possible non-thermal energies). It will be interesting to start here from a comparison among these different energies, based on radial profiles measured by different authors for the radial distribution of coronal densities, temperatures, and magnetic fields. The top panel of Fig.~\ref{Fig1} shows different radial profiles of the coronal magnetic field strengths $B(r)$ as measured with different techniques, and in particular by \citet{Dulk1978} (solid blue line), \citet{Mancuso2013} (dotted blue line), \citet{Patzold1987} (dashed blue line), \citet{Gopalswamy2011} (dashed-dotted blue line), \citet{Susino2015} (dash-triple dotted blue line). Overall these radial profiles show a quite good agreement, hence the magnitude of the coronal magnetic field is on average quite well constrained. The problem, as mentioned in the Introduction, is how to measure the magnetic fields locally, and hence how to derive their latitudinal distribution at least on the plane of sky (POS).

Starting from these magnetic field profiles, it is straightforward to derive the corresponding radial profiles of the magnetic energy density $E_m(r) = B^2(r)/2\mu$, that are shown in the bottom panel of Fig.~\ref{Fig1} (same linestyles as those in the top panel for different references). What will be interesting is also to compare these magnetic energies with other kinds of energies. In particular, starting from the coronal electron density radial profiles $n_e(r)$ it is straightforward to derive the gravitational potential energy density $E_g(r) = - G M_\odot \rho(r) /r$, where $G$ is the gravitational constant, $M_\odot$ is the solar mass, $\rho = n_e \mu_e m_H$ is the plasma mass density, $m_H$ is the Hydrogen atom mass, and $\mu_e = (1+4\alpha)/(1+2\alpha)$ with $\alpha = n_\alpha/n_p$ as the Helium abundance \citep[see e.g.][]{Lemaire2011}. The curves shown in the bottom panel of Fig.~\ref{Fig1} have been obtained by assuming $\alpha = 0.05$ \citep{Raymond1998b, Laming2003}, and the density profiles provided by \citet{Gibson1999} (solid green line), \citet{Cranmer1999} (dotted green line), \citet{Saito1977} (dashed green line), \cite{Leblanc1998} (dashed-dotted green line). Moreover, starting from the electron temperature radial profiles $T_e(r)$, these energies can be compared with the coronal plasma thermal energy density $E_t(r) = 2\,n_e(r) k_B T_e(r)$ (with $k_B$ denoting the Botzmann constant), as done here by assuming the radial density and temperature profiles provided by \citet{Gibson1999} (solid red line) and by \citet{Cranmer1999} (dotted red line). It is also easy to include the comparison to the radial distribution of the kinetic energy density $E_v(r) = 1/2\, \rho(r) v(r)^2$ associated with the fast solar wind \citep[solid orange line, obtained by assuming the density $\rho(r)$ and velocity $v(r)$ profiles by][]{Cranmer1999} and with the slow solar wind \citep[dotted orange line, obtained by assuming the density and velocity profiles by][]{Gibson1999, Bemporad2021}. Other energy sources (e.g. plasma waves, turbulence, etc...) are expected to generate energies that are lower than the energies already provided.

Overall the bottom panel of Fig.~\ref{Fig1} shows that the dominant energy density in the solar corona is related to the magnetic field, a result that is not surprising considering that the coronal dynamics are dominated by the magnetic fields. More interestingly, the bottom panel of Fig.~\ref{Fig1} also shows that among all of the types of energies, only the gravitational potential energy has values (considering the absolute value) that are comparable at all altitudes with the magnetic energy, and the plasma thermal and kinetic energies are in general much lower. Moreover, the bottom panel of Fig.~\ref{Fig1} shows that in the inner corona ($r < 5$ \rsun) the absolute value of the gravitational potential energy density is about a factor $\sim 2$ smaller than the magnetic energy density, suggesting a possible relationship between these two kinds of energies. The latter consideration is very important, and is discussed in the following section.

\subsection{Relating the magnetic and gravitational energies}

In the inner solar corona, possible relationships between the gravitational and the magnetic energy densities were explored for a long time in the previous literature, starting from the first attempts to reconstruct analytically the coronal magnetic field structure in magneto-static atmospheres \citep[e.g.][]{Low1975, Hundhausen1981, Low1982, Low1993, Hundhausen1994}. These works usually assumed a static equilibrium among the gravitational, Lorentz, and pressure forces, mostly aimed at demonstrating how density depletions and enhancements may form at the base of coronal streamers, and at deriving analytic solutions for the magnetic field distributions in the inner corona. More recently, these works were extended also to consider twisted fields in polytropic atmospheres \citep{Flyer2005}, and also solar wind expansion \citep{Sun2005}. 

\begin{figure*}[t!]
	\centering
	\includegraphics[width=0.9\textwidth]{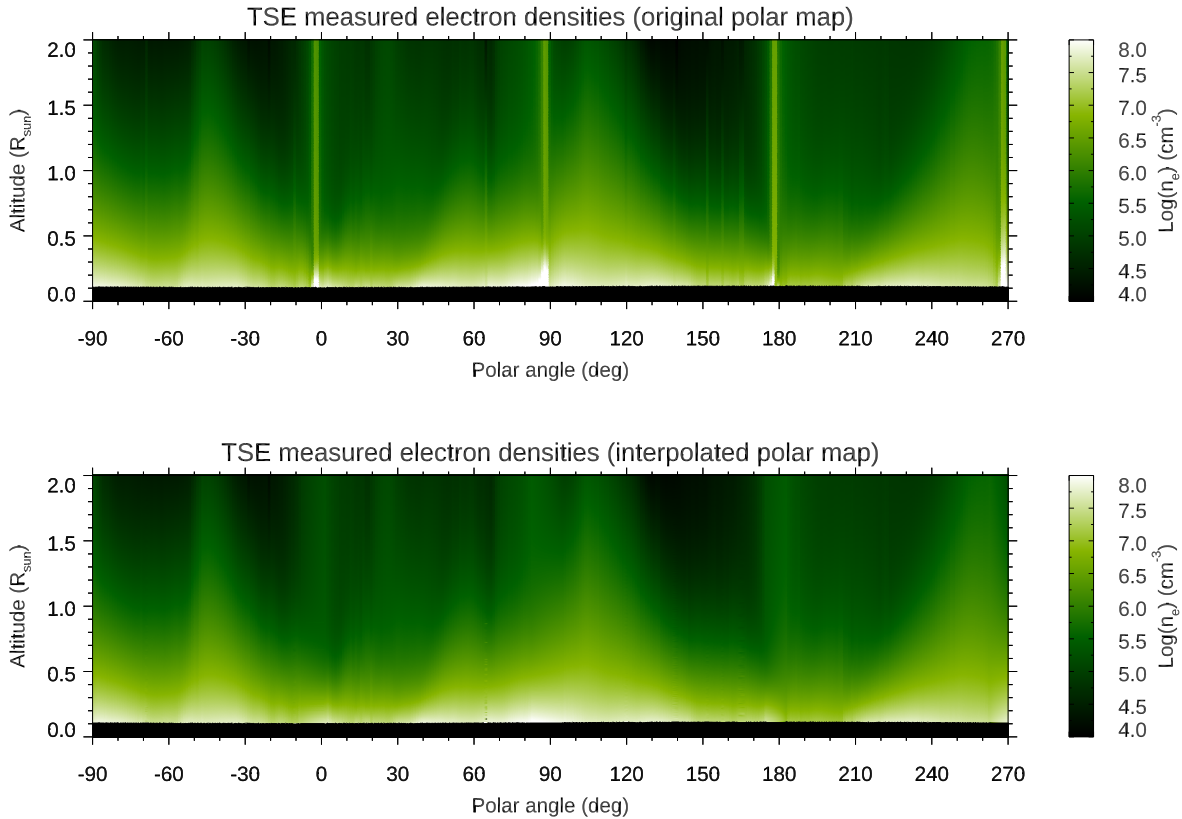}
	\caption{
		Top: Electron density map as derived from the inversion of coronal $pB$ image and converted into polar coordinates. Bottom: same as the above map, but interpolated at latitudes of missing data \citep[see][for details]{Bemporad2020}. 
	}
	\label{Fig2}
\end{figure*}
On the other hand, far from the Sun the solar wind speed and the plasma density are usually anti-correlated \citep[see][and references therein]{Richardson2003}, thus providing an almost constant solar wind mass flux. This property, recently also observed in the intermediate corona with remote sensing data \citep{Bemporad2017}, is usually ascribed to an anti-correlation near the Sun between the solar wind acceleration and the magnetic flux-tube expansion rate: if the plasma is propagating only along the magnetic flux-tubes, a super-radial expansion of their cross sections is also responsible for a larger decompression of the plasma thus reducing the plasma density by more than a factor of $1/r^2$ \citep[see e.g.][]{Wang1990}. In-situ data (acquired by the Ulysses mission) also revealed the existence of the so-called pressure-balanced structures, in which fluctuations of the magnetic and plasma pressures balance each other to maintain a comparatively constant total pressure \citep{McComas1996}. In particular \citet{Reisenfeld1999} discussed about the possible solar origin of pressure-balanced structures in polar regions in the plume-interplume interactions, suggesting that going from the inner regions, where the plasma $\beta \ll 1$ to the outer regions, where $\beta > 1$, in the plume regions with super-radial expansion the plasma is decompressed, while the opposite occurs in the interplume regions, where the plasma could be compressed. Hence, a similar effect could be responsible in the corona for an anti-correlation between the local density (and hence the gravitational energy density) and the magnetic field strength (and hence the magnetic energy density).

\begin{figure*}[t!]
	\centering
	\includegraphics[width=\textwidth]{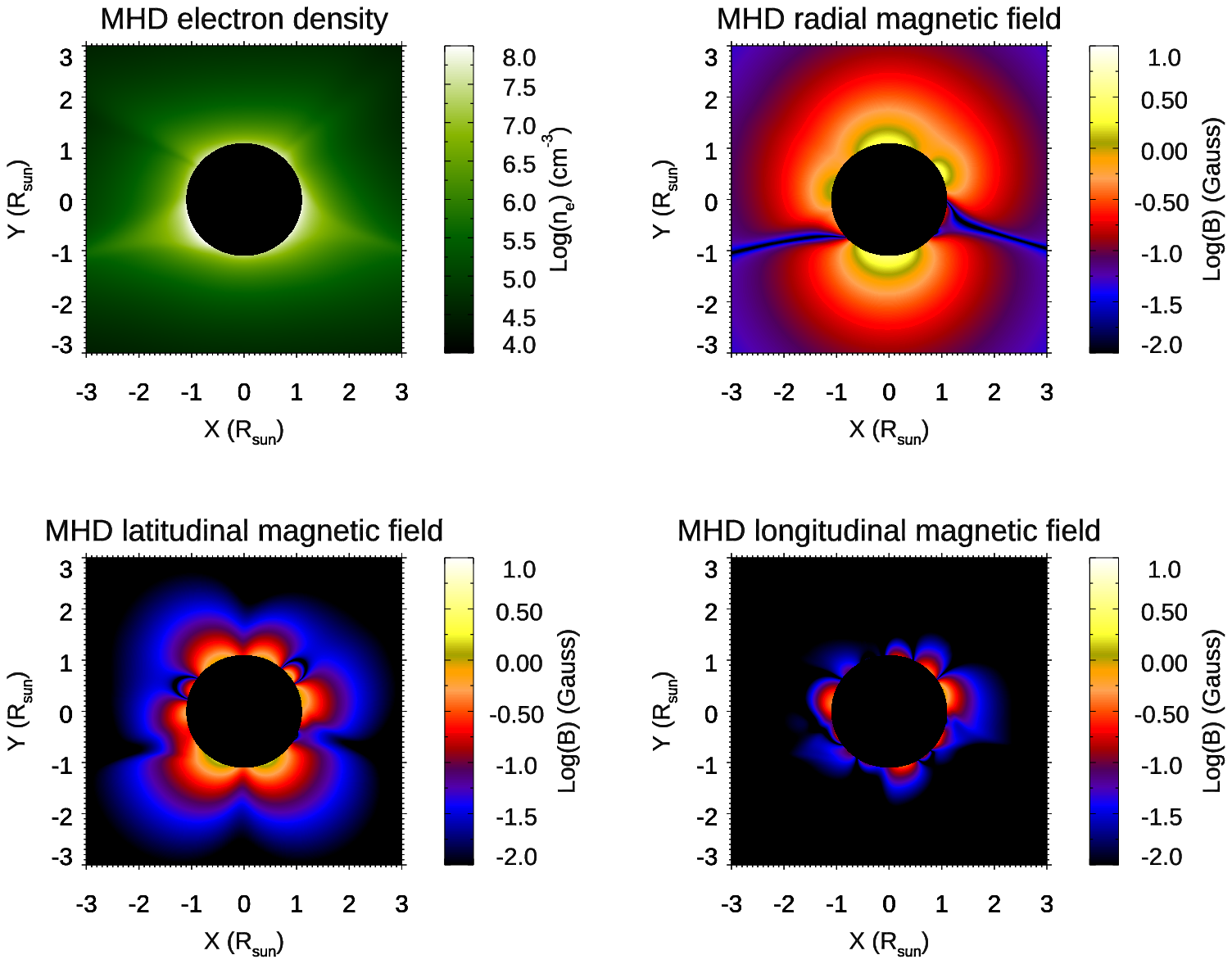}
	\caption{
		Plasma physical parameters predicted for the 2017 TSE by the MHD numerical simulations performed with the MAS model. Different panels show the POS distribution of coronal electron densities (top left), and the absolute values of the radial (top right), latitudinal (bottom left) and longitudinal (bottom right) components of the coronal magnetic fields. 
	}
	\label{Fig3}
\end{figure*}
More recently, the existence of an anti-correlation between the coronal electron densities and magnetic field strength was noticed by \citet{Susino2015}, but this anti-correlation was mostly ascribed to the technique that was developed in that work to measure the magnetic field from the propagation of MHD shocks associated with Coronal Mass Ejections. In a previous work, \citet{Dolei2014} demonstrated that the motion of chromospheric plasma blobs ejected into the solar corona after the spectacular 7 July 2011 solar eruption is affected not only by the gravitational force, but also by a magnetic drag force, almost comparable in strength with the gravitational force. In the interpretation proposed by \citet{Dolei2014}, adapted from that proposed by \citet{Haerendel2011} for prominences, this magnetic drag force results from the interaction between the moving plasma blob and the background coronal magnetic field. The ballistic motion of the blob perturbs the crossed magnetic field exciting \alfven\ waves, causing a net transfer of energy from the mechanical energy possessed by the blob to the energy of the propagating \alfven\ waves \citep[see][Figure 5]{Dolei2014}. A very interesting result from this analysis is that by measuring the mechanical energy lost by the blob as a function of height, and by assuming that this is entirely due to the excitation of Alfv\'en waves, the resulting magnetic field values are very close to the reference curve provided by \citet{Dulk1978}. Hence, the potential energy loss seems to be proportional to the energy transferred to the magnetic field. Therefore, if a blob reaches a certain height $h$ with instant speed 0 just before falling back to the Sun, at that moment the embedded plasma has a potential energy (apart from the thermal and magnetic energy of the plasma blob) which is lower than the total mechanical energy it possessed at the moment of expulsion, due to magnetic drag force that transferred part of the mechanical energy to outwardly propagating \alfven\ waves. These considerations seem to suggest again a possible relationship in the coronal plasma between the gravitational potential energy and the energy of the background magnetic field.

\subsection{Energy equipartition in the solar corona}

A basic principle that could be employed to relate different kinds of energies is the equipartition principle. In general, an equipartition principle states that the total energy of the system is equally distributed among all of its components (magnetic, kinetic, thermal, etc...). Usually, for plasma with $\beta \ll 1$ (with $\beta = p_{gas}/p_{mag}$, $p_{gas} = 2n_e k_BT_e$ and $p_{mag} = B^2/2\mu$) this principle is applied by considering that the dominant energies in the plasma are the magnetic energy $E_m$ and the kinetic energy $E_v$, hence
\begin{equation}
	\frac{B^2}{2 \mu} = \frac{1}{2} \rho v^2.
\end{equation}
Values of the magnetic field (or at least a lower limit of it, considering other energies being neglected) can be derived from the above simple relationship, once the plasma density $\rho$ and velocity $v$ are known. This equality has been assumed for the measurement of magnetic fields in the solar photosphere \citep[e.g.][]{Beckers1971}, and in prominences from the observations of moving blobs \citep[e.g.][]{Ballester1984} or knots \citep{Zapior2012, Zapior2016}, hence in regimes where $\beta > 1$. An equipartition between magnetic and kinetic energy densities also holds for \alfven\ waves \citep[e.g.][]{Cranmer2005} in the corona. However, departures from the equipartition principle have been found for small-scale coronal eruptive events like polar jets \citep{Paraschiv2015}, and in chromospheric fibrils \citep{Parker1984}. Laboratory plasma experiments and theoretical considerations suggest that during the fundamental process of magnetic reconnection an equipartition principle between magnetic and kinetic energies cannot be applied \citep[see discussion in][]{Bemporad2008}. Hence, these previous works suggest that for dynamic processes the equipartition principle cannot be applied.

On the other hand, considering the large-scale stationary solar corona as shown in Fig.~\ref{Fig1} it appears that the dominant energies are not the magnetic and kinetic energies, but the magnetic and gravitational potential energies. Hence, if the equipartition principle is considered for the large-scale stationary solar corona, this should relate these two latter major kinds of energies. All these considerations suggest in the end the idea that a relationship could exist between the coronal plasma potential energy distribution $E_g$ and the magnetic field energy $E_m$, a possible relationship that I will investigate in the next paragraphs.

\section{Input observations and modeling} \label{sec: input}

\subsection{Coronal observations}

This work is based on the electron densities measured in the solar corona in the occasion of TSE that occurred on August 21st 2017 and visible from across the United States. A detailed description of the images acquired and of the data analysis methods is provided in \citet{Bemporad2020}. Here I simply remind the reader that during these observations the polarized images were acquired only with two different orientations of a linear polarizer, separated by about 90$^\circ$. For this reason, the derived polarized brightness (pB) image (and hence the corresponding electron density map) diverges in coronal regions separated by 90$^\circ$ in latitude.

The electron density map $n_{TSE}$ derived in \citet{Bemporad2020} from TSE observations is provided here in the top panel of Fig.~\ref{Fig2}, after conversion from Cartesian to polar coordinates. In particular, the conversion is performed preserving the same spatial pixel size in the radial direction of the original image (2732 km pixel$^{-1}$) and with an angular resolution by 0.075$^\circ$ pixel$^{-1}$. The polar angle (PA) is measured as usual starting from the solar north pole and running clockwise, hence in this map PA = 0$^\circ$ corresponds to the North pole, while PA = 180$^\circ$ corresponds to the South pole. This map clearly shows the locations of main coronal streamers observed on that day, as well as a few limited latitudinal regions where a reliable measurement of the density was not derived. In order to reduce the effect of these missing data, the density map is interpolated here in these latitudinal regions (having angular extensions of $\sim 8^\circ - 10^\circ$) with no reliable measurements. The resulting interpolated polar map is provided in the bottom panel of Fig.~\ref{Fig2}. This density map is the starting point for the analysis described in the rest of this work and for the derivation of coronal magnetic fields.
\begin{figure*}[t!]
	\centering
	\includegraphics[width=0.9\textwidth]{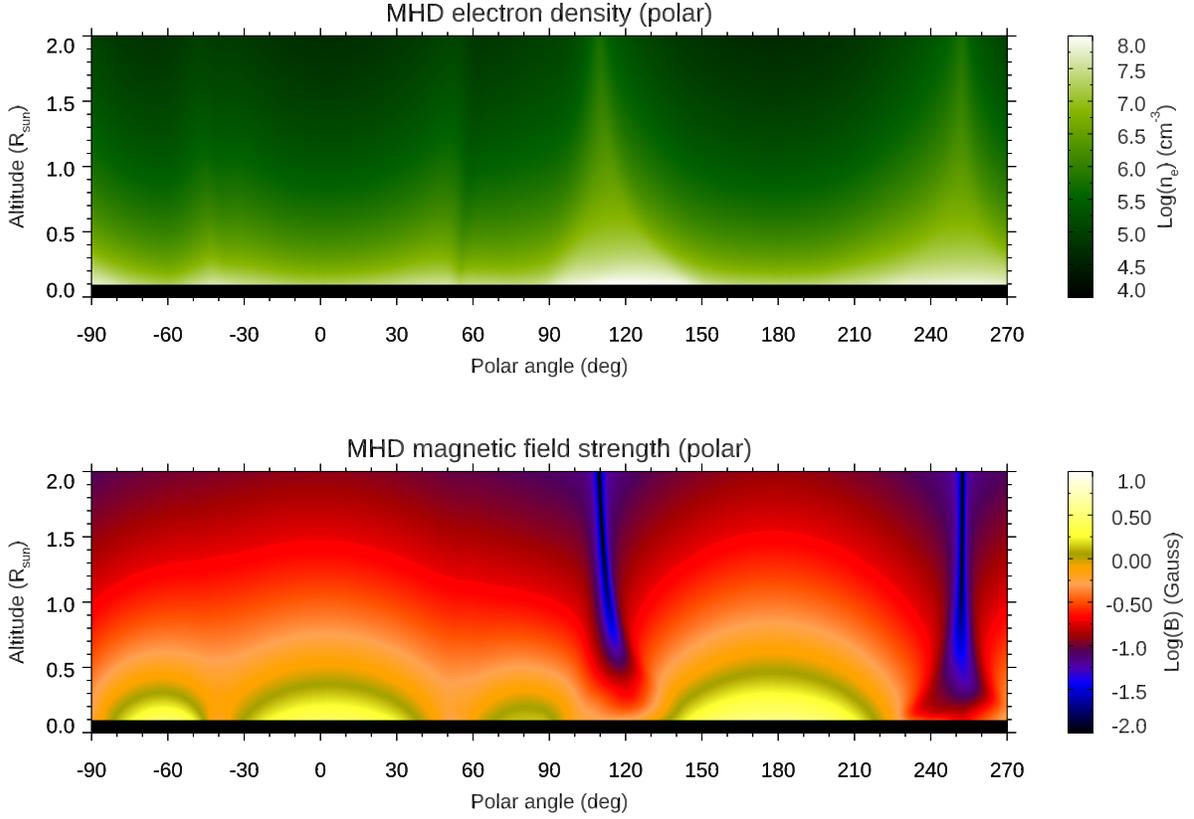}
	\caption{
		Polar maps of the POS coronal electron densities (top) and magnetic field strength (bottom) as predicted by the MHD numerical simulations performed with the MAS model for the 2017 TSE. 
	}
	\label{Fig4}
\end{figure*}

\subsection{Coronal modeling}

At present, there are no routine measurements of the global coronal magnetic fields. In any case, a reliable reference for this quantity is usually provided by the MHD numerical simulations reconstructing these fields from the photospheric field measurements acquired over one full solar rotation. One of the most advanced and currently available models reconstructing the global coronal magnetic fields is the MHD algorithm outside a sphere (MAS) code \citep[e.g.][]{Linker1999, Mikic1999}, and in particular the MAS 3D MHD thermodynamic code, which also includes coronal heating, anisotropic thermal conduction, and radiative losses, thus providing one of the more realistic simulations of the global plasma currently available \citep{Lionello2009}. Hence, in this work I assume as a reference the plasma physical parameters (in particular, the density and magnetic field) reconstructed with this model for the 2017 TSE on the POS. The reason why in this work the POS values from the model are selected for a comparison with the LOS-integrated values from the observations is to show that, despite the existing LOS-integration effects, the resulting plasma parameters derived from the observations are in very good agreement with those predicted on the POS by the MHD simulations, hence the LOS integration effects can be neglected.

\begin{figure*}[!t]
	\centering
	\includegraphics[trim={1cm 2cm 2cm 0}, width=0.95\textwidth]{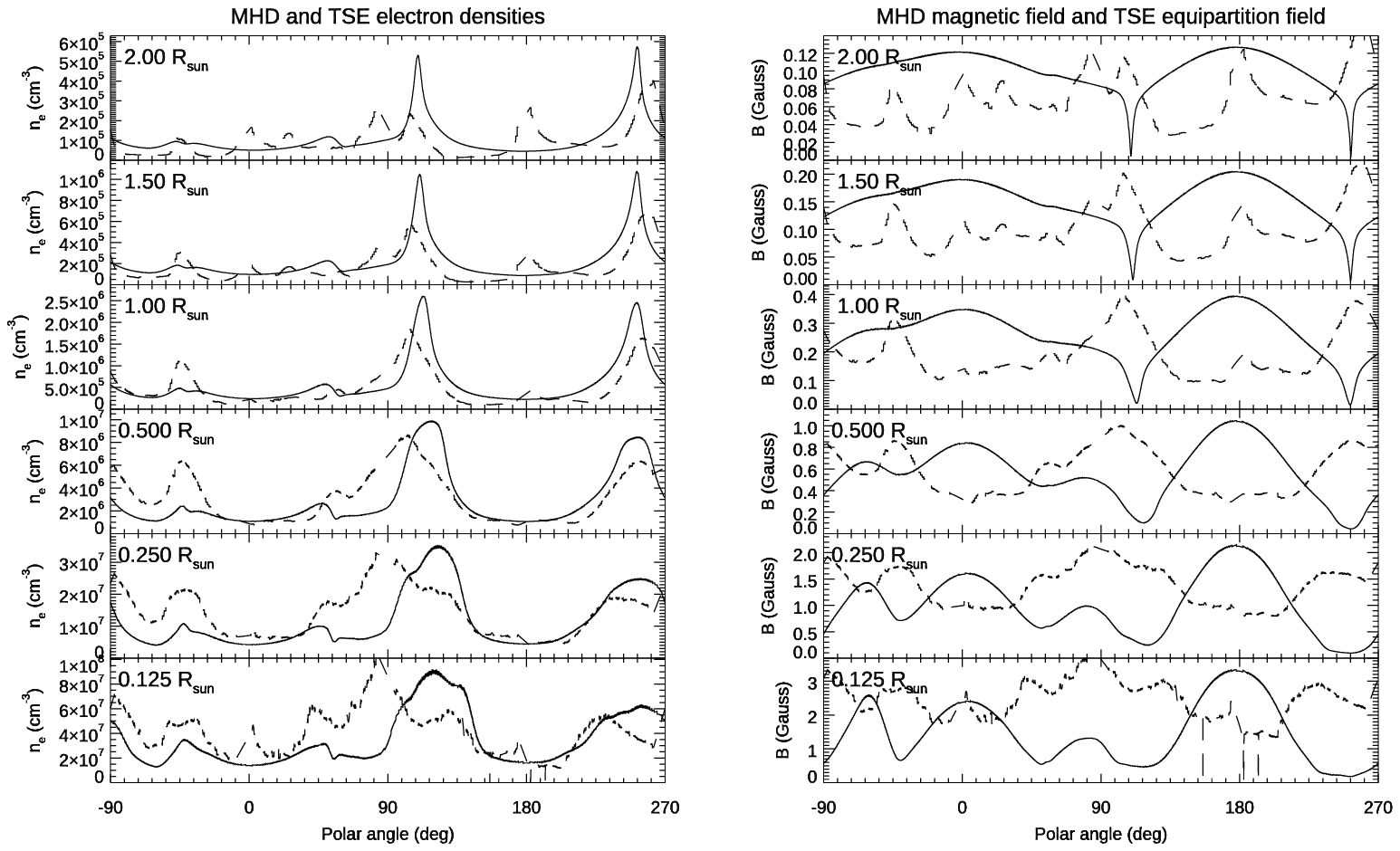}
%	\vspace{-5cm}
	\caption{
		Left column: latitudinal distributions at different heights above the solar surface of the coronal electron densities predicted by the MHD numerical simulations (solid lines) and measured by TSE observations (dashed lines). Right column: latitudinal distributions of coronal magnetic field strength predicted by the MHD numerical simulations (solid lines) and the magnetic field values provided by the equipartition hypothesis from TSE measurements (dashed lines, see text for explanations).
	}
	\label{Fig5}
\end{figure*}

These two plasma parameters are shown in Fig.~\ref{Fig3}. In particular, this Figure provides the POS distributions of the coronal electron density $n_{MHD}$ (top left), and the absolute values of the three components of the coronal magnetic fields $B_{MHD}$ from MHD reconstruction: the radial $B_r$ (top right), latitudinal $B_\theta$ (bottom left), and longitudinal $B_\phi$ (bottom right) components. The three magnetic field components are provided with the same Log scale, to show that - as expected - $B_r$ is dominant, and only at the inner altitudes $B_\theta$ and $B_\phi$ components are contributing significantly to the total magnetic field strength. It is interesting to point out here that the two brightest coronal streamers shown in the density map (top left in Fig.~\ref{Fig3}) above the southern hemisphere, correspond in the $B_r$ map (top right in Fig.~\ref{Fig3}) to two "black" features (extending almost radially above the solar surface) corresponding to the absolute minima of the map. These features correspond to the latitudinal locations where the streamer current sheets are located, and hence the radial component of the magnetic field $B_r$ changes signs, passing through zero.

\subsection{Comparison between observations and modeling}

In order to facilitate the comparison with the polar density map from TSE observations provided in Fig.~\ref{Fig2}, the electron number density $n_{MHD}$ and the total field strength $B_{MHD} = \sqrt{B_r^2+B_\theta^2+B_\phi^2}$ from the MHD simulation are also provided in polar coordinates, and are shown respectively in the top and bottom panels of Fig.~\ref{Fig4}. A comparison between the top map of Fig.~\ref{Fig4} and the bottom map of Fig.~\ref{Fig2} shows the overall agreement between the predicted $n_{MHD}$ and the observed $n_{TSE}$ coronal densities, with the two large coronal streamers located above the South hemisphere and centered at PA $\simeq 110^\circ$ (latitude of $20^\circ$ SE) and PA $\simeq 250^\circ$ (latitude of $20^\circ$ SW), and the two smaller radial features observed above the North hemisphere and centered at PA $\sim -50^\circ$ (latitude of $40^\circ$ NW) and PA $\simeq 60^\circ$ (latitude of $30^\circ$ NE). A notable disagreement between observations and simulations is related to the pseudostreamer located above the Northwest limb which is not correctly reproduced by the simulations, mainly because of inaccuracies in the synoptic photospheric magnetic field data, resulting in inaccuracies in the reconstructed coronal magnetic fields \citep[see discussion by][for more details]{Mikic2018}.
\begin{figure*}[!t]
	\centering
	\includegraphics[width=0.7\textwidth]{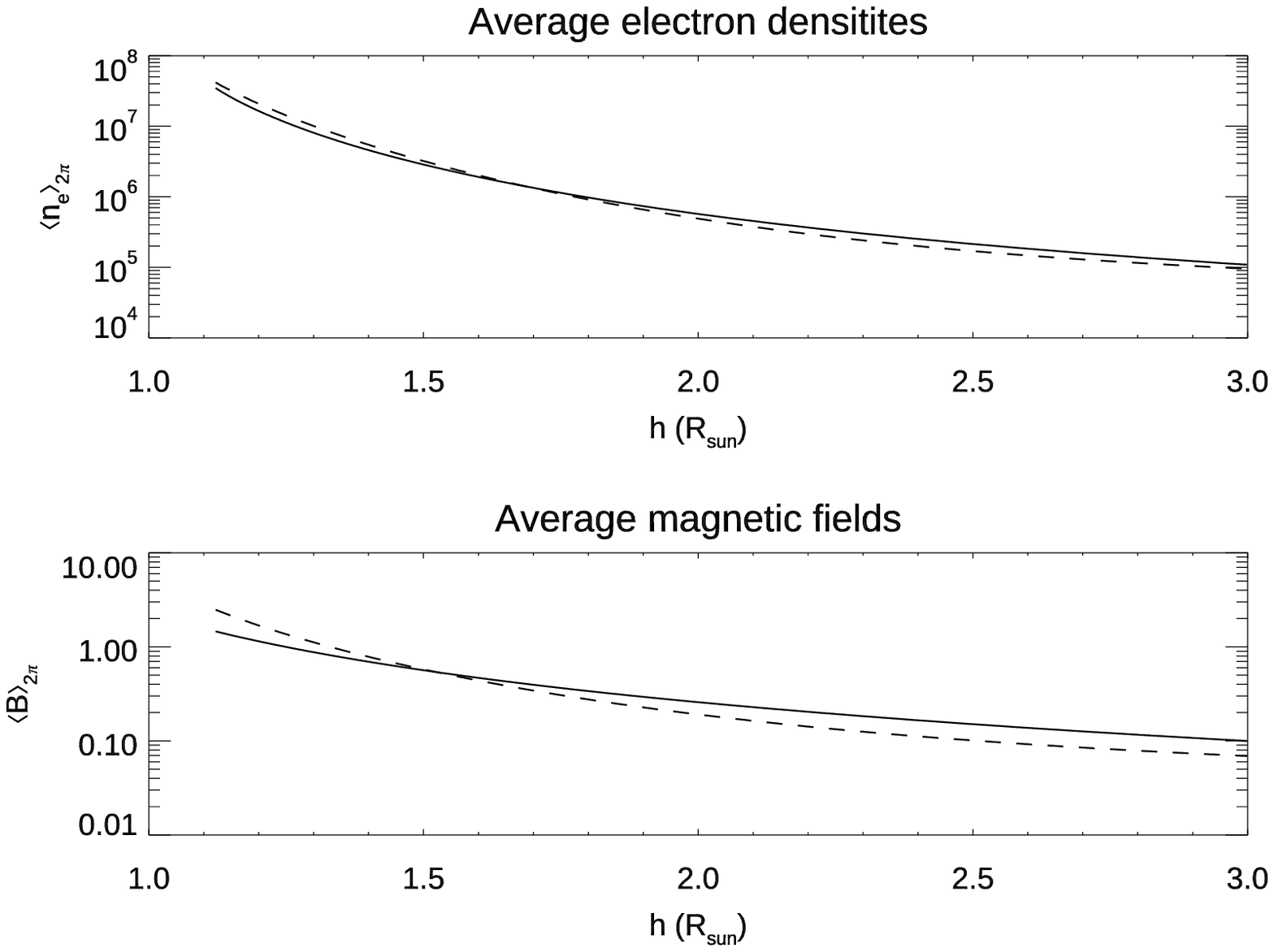}
%	\vspace{-2.8cm}
	\caption{
		Radial distribution of coronal electron densities ($\langle n_e \rangle_{2 \pi}$, top panel) and magnetic fields ($\langle B \rangle_{2 \pi}$, bottom panel) averaged over all latitudes. Both panels show the values predicted by the MHD numerical simulations (solid lines) and measured by TSE observations (dashed lines).
	}
	\label{Fig6}
\end{figure*}

The two southward features are characterized by the inversion of the radial magnetic field, and hence by the presence of a current sheet, and can thus be classified as classical coronal streamers; the same does not apply to the two northward features, which cannot be classified as proper coronal streamers. Apparently, this information cannot be derived only from the observed coronal density maps, and necessarily requires the knowledge of coronal magnetic fields reconstructed with MHD simulations. On the contrary however, as I will show in the rest of this work, this information can be derived from the measured coronal densities alone.

\section{Coronal magnetic field strength} \label{sec: field strength}

\subsection{Equipartition magnetic field}

As briefly discussed in the introduction, the dominant energies in the solar corona appear to be the gravitational potential energy $E_g$ and the magnetic energy $E_m$. Starting from this consideration, one can test a different version of the equipartition principle in which the total magnetic energy is equally distributed between the potential energy on one side, and all the other energy forms on the other side (including kinetic energy, thermal energy, etc...), hence
\begin{equation}
    \frac{1}{2}\frac{B^2}{2\mu} = \frac{G\, M_\odot\, \rho}{r}. \label{eq: equipartition principle}
\end{equation}
Notice that the above expression is in agreement with the finding previously obtained that on average in the solar corona the dominant energy is related to the magnetic field and that the gravitational potential energy is about a factor $\sim 2$ lower at all altitudes (Fig.~\ref{Fig1}). The assumption that in the solar corona the dominant energies are $E_g$ and $E_m$ provides the definition of the "equipartition magnetic field" $B_{eq}(r,\theta)$ which is given by
\begin{equation}
    B_{eq}(r,\theta) = \left( \frac{4\mu \,G\, M_\odot\, \rho(r,\theta)}{r} \right)^{1/2}. \label{eq: equipartition field}
\end{equation}
This simple relationship can be employed as a test to convert a 2D map of the coronal electron densities as a function of height $r$ and latitude $\theta$ into a 2D map of the equipartition magnetic field as defined above.

The results from this test, once the above relationship is applied to the polar density map derived from 2017 TSE observations, are given in Fig.~\ref{Fig5}. First of all, the left panels of Fig.~\ref{Fig5} provide a comparison between the latitudinal distributions of the measured coronal number densities ($n_{TSE}$, dashed lines) and those predicted by the MHD simulations ($n_{MHD}$, solid lines) at different altitudes going from $r = 0.125$ \rsun\ (bottom panel) up to $r = 2.0$ \rsun\ (top panel). This comparison shows overall a quite good agreement between $n_{TSE}$ and $n_{MHD}$, suggesting that the corona reconstructed with the MHD simulation is overall well reproduced in most coronal regions. On the other hand, a more detailed comparison also shows some differences, in particular in the inner corona where the real observations show that the South-East streamer extends towards the equatorial region much more than what is predicted, while a better agreement is found for the orientation of the South-West coronal streamer. Also, the location of the North-West coronal feature is well predicted, but the predicted densities $n_{MHD}$ are underestimated by about a factor 2 with respect to the measured densities $n_{TSE}$ \citep[see also the discussion by][]{Mikic2018}.

More interestingly, the right panels of Fig.~\ref{Fig5} show a comparison between the latitudinal distributions at different heights of the total magnetic field strength $B_{MHD}$ predicted by the MHD simulation (solid lines), and the values of the equipartition magnetic field $B_{eq}$ (dashed lines). Unexpectedly, the comparison in the right panels of Fig.~\ref{Fig5} shows at least two interesting results. First, at each altitude the average value of $B_{eq}$ is approximately of the same order of $B_{MHD}$. This is really surprising, considering that no information related to the photospheric magnetic fields (which are employed by the MHD simulation to reconstruct the coronal fields) is present in the derived plasma density map. Second, the latitudinal distribution of $B_{eq}$ appears to be at all altitudes anti-correlated to the latitudinal distribution of $B_{MHD}$. These two very interesting results (never been reported so far) are the starting point for the definition of the "equipotential magnetic field" in the next Section.
\begin{figure}
	\centering
	\includegraphics[trim={0.5cm 2cm 2cm 0.5cm}, width=\textwidth]{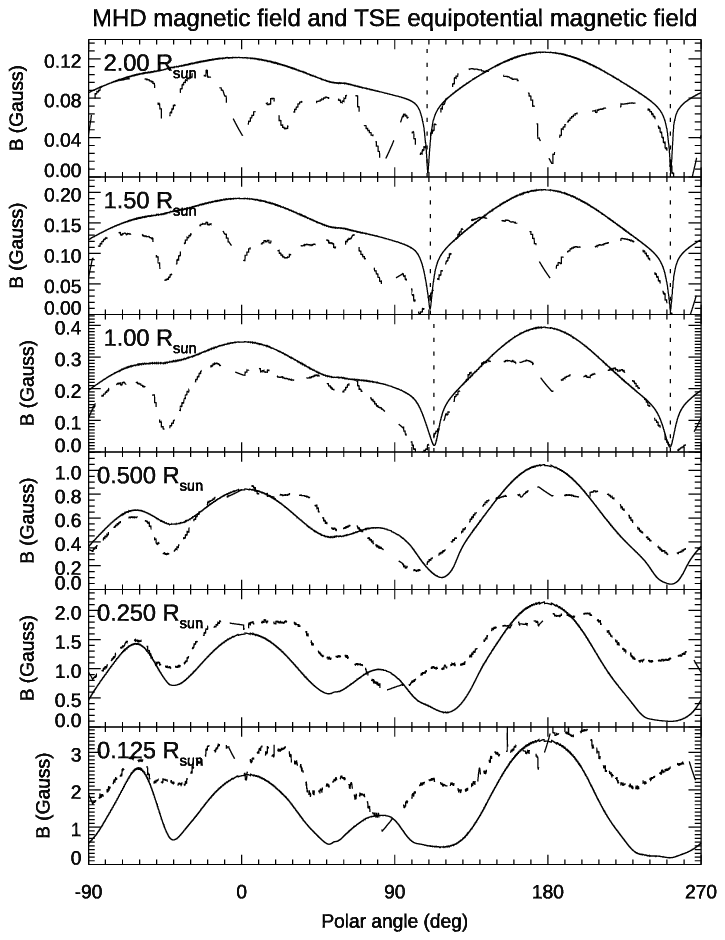}
%	\vspace{-2.8cm}
	\caption{
		Latitudinal distributions at different heights above the solar surface of the coronal magnetic field $B_{MHD}$ strength predicted by the MHD numerical simulations (solid lines) and the values of the magnetic field $B_{TSE}$ as derived from TSE measurements (dashed lines, see text for explanations). The vertical dotted lines show the latitudes of the streamer current sheets.
	}
	\label{Fig7}
\end{figure}

\subsection{Equipotential magnetic field}

The results mentioned above for the equipartition magnetic field $B_{eq}$ (Eq.~\ref{eq: equipartition field}) suggest to define a revised estimate of the coronal magnetic field, called equipotential magnetic field, as follows:
\begin{equation}
    B_{pot}(r,\theta) = 2\,\langle B_{eq}(r, \theta) \rangle_{2 \pi} - B_{eq}(r, \theta), \label{eq: equipotential field}
\end{equation}
where the symbol $\langle ... \rangle_{2 \pi}$ denotes the average obtaind at a constant altitude $r$ over the entire available latitudinal interval $2\pi$. The above simple expression allows to rescale the equipartition magnetic field, converting the observed anti-correlation with the predicted magnetic field into a real correlation and preserving the average value at all latitudes (in fact $\langle B_{pot} \rangle_{2 \pi} \equiv \langle B_{eq} \rangle_{2 \pi}$). The values for the coronal electron density and magnetic field radial profiles averaged over all latitudes are shown in Fig.~\ref{Fig6}, in the top and bottom panels, respectively. This Figure shows very good agreement between the values predicted by the MHD numerical simulation (solid lines) and those derived from TSE observation (dashed lines), not only between the electron densities $\langle n_e \rangle_{2 \pi}$, but also between the magnetic fields $\langle B \rangle_{2 \pi}$ derived with the above definition of $B_{pot}$.

The magnetic field values $B_{pot}$ resulting at all latitudes from the above semi-empirical expression are shown in the panels of Fig.~\ref{Fig7} again at different altitudes from $r = 0.125$ \rsun\ (bottom panel) up to $r = 2.0$ \rsun\ (top panel). Each plot shows in particular the latitudinal distribution of $B_{pot}$ (dashed line) in comparison with the latitudinal distribution of $B_{MHD}$ (solid line). The plots in Fig.~\ref{Fig7} show at least two main results: first of all, there is very good agreement between the $B_{pot}$ and $B_{MHD}$ values at all latitudes and all altitudes. Second, the latitudinal distribution of $B_{pot}$ is also characterized by the presence of two absolute minimum regions in correspondence to the two coronal streamers, and hence in the locations of their current sheets (vertical dotted lines in Fig.~\ref{Fig7}). Hereafter I will indicate as $B_{TSE}$ the coronal magnetic field estimated from the TSE coronal densities with the above formula for $B_{pot}$.
\begin{figure*}[t]
	\centering
	\includegraphics[width=0.9\textwidth]{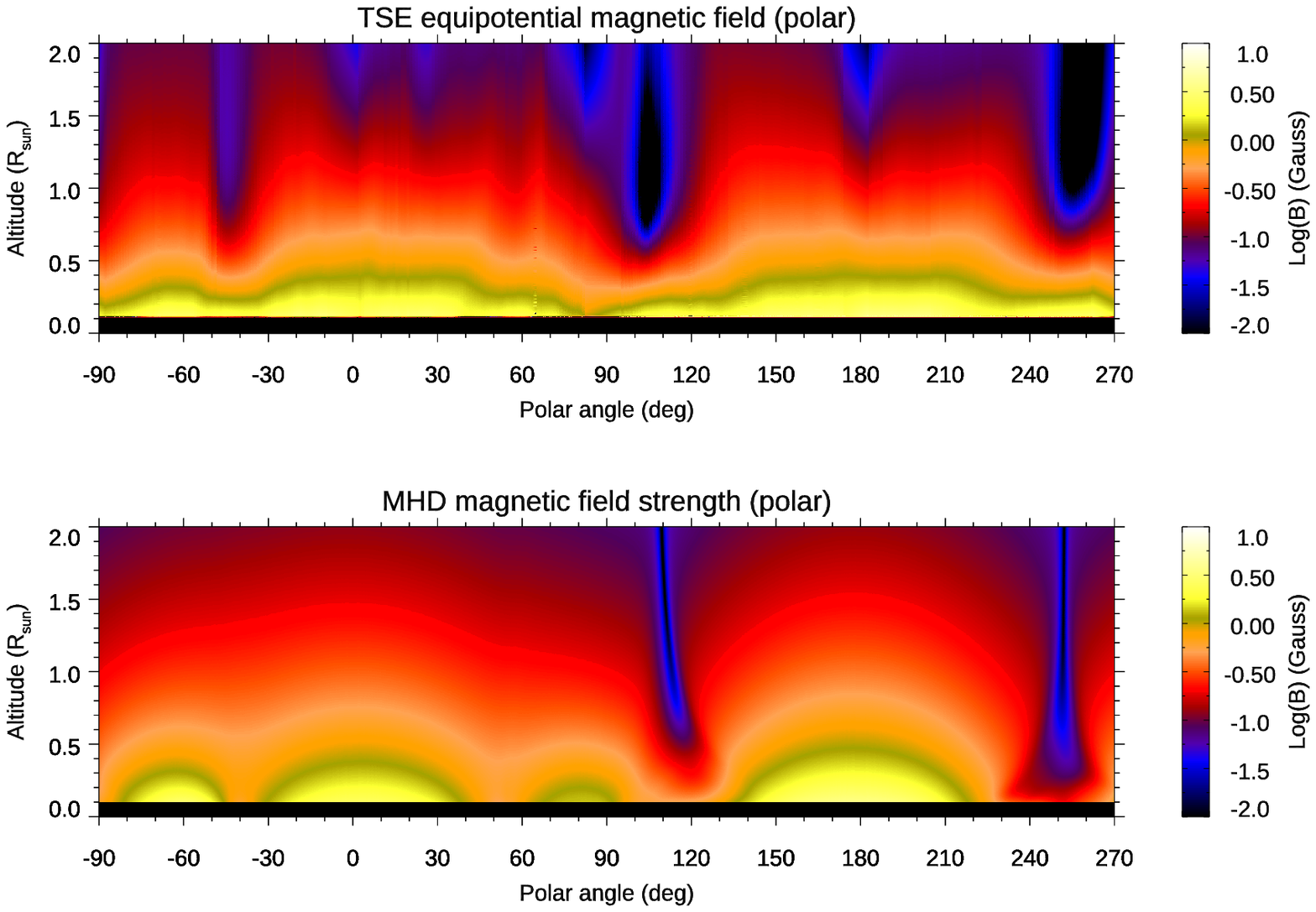}
	\caption{
		Top: polar map of the equipotential coronal magnetic field strength as derived from TSE measurements. Bottom: corresponding polar map of the coronal magnetic field strength predicted by the MHD numerical simulations.  
	}
	\label{Fig8}
\end{figure*}

The striking similarity between the $B_{TSE}$ and $B_{MHD}$ values at all latitudes and altitudes is also demonstrated by the two polar maps provided in Fig.~\ref{Fig8} in the same Log scale. The comparison between these two maps also shows many differences, but considering that (as mentioned above) the MHD simulations do not perfectly reproduce the observed coronal density distribution, and hence the coronal plasma conditions, it is very hard to understand at this level if the differences are due to limitations of the $B_{MHD}$ reconstruction, or to limitations of the $B_{TSE}$ computation. These two polar maps are also provided in Cartesian coordinates in the left panels of Fig.~\ref{Fig9}, again with the same Log scale. These maps show again not only the overall agreement between the two magnetic field estimates, but also the clear identification made possible by the derivation of $B_{TSE}$ of the current sheets crossing the coronal streamers.

\subsection{Other plasma parameters}

\begin{figure*}[t!]
	\centering
	\includegraphics[width=\textwidth]{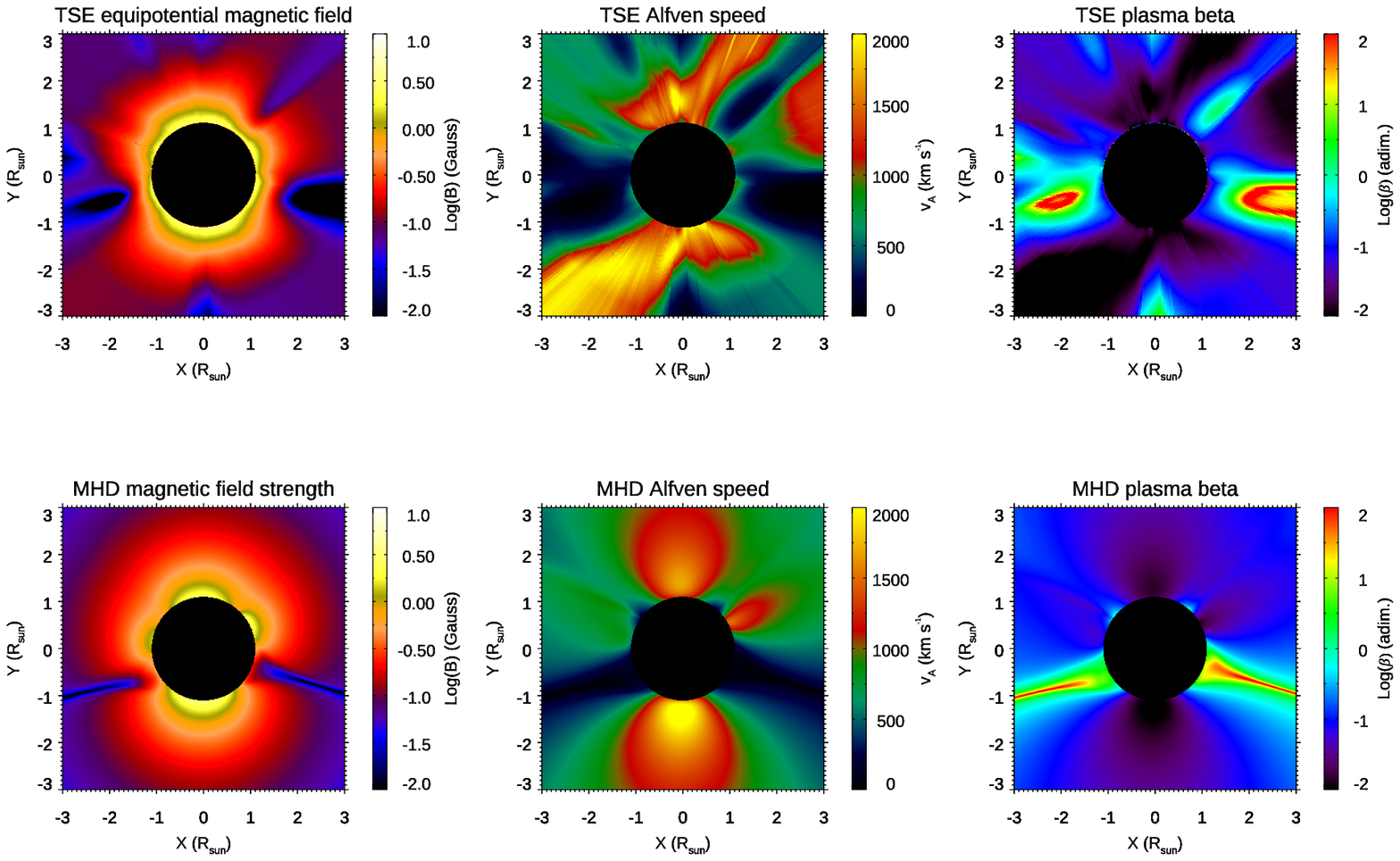}
	\caption{
		Top panels: POS distributions of the coronal magnetic field strength (left),  Alfv\'en speed (middle) and plasma$-\beta$ (right) as derived from TSE measurements. Bottom panels: the corresponding quantities as predicted by the MHD numerical simulations.
	}
	\label{Fig9}
\end{figure*}

Given the 2D distributions of coronal magnetic fields $B_{TSE}$ and of coronal mass densities $\rho$, it is straightforward also to derive the 2D distribution of the local Alfv\'en speed $v_A$ which is given by
\begin{equation}
    v_A (r,\theta) = \frac{B_{TSE}(r,\theta)}{\sqrt{\mu\, \rho(r,\theta)}}.
\end{equation}
The middle panels of Fig.~\ref{Fig9} show (with the same linear color scale) the comparison between the POS distribution of $v_A$ from the MHD numerical model (bottom middle) and the one derived with the above formula from TSE observations (top middle). Again, the similarities between the two maps are evident, demonstrating that the above derivation of $B_{TSE}$ also allows to derive a good estimate for the 2D distribution of the Alfv\'en speed in the inner corona. The derivation of the $v_A$ distribution in the solar corona is of fundamental importance for instance to understand where coronal shocks form during the propagation of solar eruptions, and hence where and when energetic particles can be more efficiently accelerated by these shocks. 

More than that, the coronal magnetic field and electron density measurements derived with TSE observations can be combined with possible values of the coronal electron temperatures $T_e$ to infer the POS distribution of the so-called plasma$-\beta$ parameter defined as
\begin{equation}
    \beta = \frac{2\,n_e k_B T_e}{B^2/2\mu}.
\end{equation}
As is well known, this (adimensional) parameter, given by the ratio between the plasma gas pressure and the magnetic pressure, represents the relative importance of magnetic forces in the plasma dynamic phenomena. From the observational point of view the data analyzed in this work and acquired during the 2017 TSE does not allow to measure the coronal temperatures, unless hydrostatic equilibrium is assumed \citep[see e.g.][]{Gibson1999} and the hydrodynamic expansion of the solar wind is also considered \citep{Lemaire2016}. Reference values for the coronal electron temperature radial profiles $T_e(r)$ at different heliocentric distances $r$ are provided for instance by \citet{Gibson1999} and \citet{Cranmer1999} respectively for solar minimum coronal streamers and coronal holes.

Here, given the similarity between these two radial profiles, and in a similar way to what was recently done by \citet{Bemporad2021}, these profiles are averaged to derive a 2D map of coronal electron temperatures at all latitudes and at different distances $r$. The $T_e(r)$ map is then combined with $n_e(r,\theta)$ and $B_{TSE}(r,\theta)$ measurements from TSE observations to derive a 2D map of the $\beta_{TSE}$ parameter as defined above. The same quantities from MHD numerical simulation are combined to derive a 2D map of the predicted plasma $\beta_{MHD}$ POS distribution. The right panels of Fig.~\ref{Fig9} show (with the same Log color scale) the comparison between the POS distributions of $\beta_{MHD}$ (bottom right) and $\beta_{TSE}$ (top right). Both maps show that, as expected, $\beta \lesssim 1$ almost at all latitudes, with the exception of those for the two coronal streamers located in the southward hemisphere, where near the crossing of the current sheet the magnetic pressure drops almost to zero, leading to $\beta \gtrsim 1$. The overall agreement between $\beta_{MHD}$ and $\beta_{TSE}$ is also related to the fact that, among the different plasma parameters in the expression for $\beta$, the electron temperature $T_e$ has a much smaller radial gradient with respect to the electron density $n_e$ and to the magnetic field squared $B^2$. Hence, a good estimate of $\beta$ can be provided with the $n_e$ and $B_{TSE}$ values from TSE observations just by assuming an average radial profile for $T_e$, as shown above.

Before concluding this part, it is important to point out that the coronal densities employed in this work were derived in \citet{Bemporad2020} with the standard van de Hulst inversion technique \citep{Hulst1950}, and thus were affected by possible LOS-integration effects. For instance, if we consider a coronal streamer with fixed local density, by changing the inclination of the Current Sheet (CS) in the streamer belt with respect to the LOS, the density measured with this technique will be more accurately estimated if the CS is more parallel to the LOS, because the inversion assumes spherical symmetry, and it will be more likely under-estimated if the CS is more perpendicular to the LOS. This effect was found and quantified for instance by \citet{Wang2014}, who validated this technique with synthetic data and tomographic reconstructions. The opposite likely will happen for coronal hole regions: at the latitudes where coronal features (like streamers or plumes) located far from the POS are integrated along the LOS, the density measured with the standard inversion assuming spherical symmetry will be likely over-estimated with respect to the actual local electron density in the coronal hole. An over-estimate (under-estimate) of the coronal densities will lead in principle to an over-estimate (under-estimate) of the coronal equipartition magnetic field strength estimated with the method proposed here (Eq. \ref{eq: equipartition field}) as a function of latitude. Nevertheless, the values of the equipotential magnetic field (Eq. \ref{eq: equipotential field}) comes from an average of the equipartition magnetic field at all latitudes in the visible corona, hence regions with over-estimated densities (coronal holes) will counter-balance regions with under-estimated densities (coronal streamers), thus reducing possible errors.

The next Section will discuss the possible theoretical justifications for the above results.

\section{Possible theoretical justifications} \label{sec: theoretical}

The previous Sections demonstrate that a good estimate of the coronal magnetic field strength can be derived directly from the observed plasma density distribution. This Section is devoted to discussing possible theoretical justifications for the results presented here, and in particular for Eq.~\ref{eq: equipotential field}. The main difficulty in theoretically deriving Eq.~\ref{eq: equipotential field} is that this expression contains a relationship between two quantities measured locally in the corona ($B_{pot}(r,\theta)$ and $B_{eq}(r,\theta)$) and a quantity averaged at fixed altitude at all latitudes in the visible corona ($\langle B_{eq}\rangle_{2\pi}$). Following \citet{Low1993}, the starting point of this Section is the MHD virial theorem, which is extended here to include the Lorentz force and the presence of \alfven\ waves.

\subsection{Virial theorem}

The search for a first possible explanation for Eq.~\ref{eq: equipotential field} relating coronal magnetic field energy and gravitational potential energy can start from the MHD virial theorem, expressing a general relationship among different kinds of energy. In particular, the scalar MHD virial theorem \citep{Somov2007} states that for a plasma in a steady state
\begin{eqnarray}
	\int_V \left( \rho v^2+\,3p\,+\,\frac{B^2}{2\mu}\,-\,\frac{G\,M_{\odot}\rho}{r}\right)\,dV-\oint_S p\left( \mathbf{r} \cdot \mathbf{dS} \right)\,+ \label{eq: virial} \\
	- \oint_S \left[ \frac{B^2}{2\mu} \left( \mathbf{r} \cdot \mathbf{dS} \right) - \frac{1}{\mu} \left( \mathbf{B} \cdot \mathbf{r} \right) \left( \mathbf{B} \cdot \mathbf{dS} \right) \right] = 0 \nonumber
\end{eqnarray}
where $\int 3p\, dV = 3 (\gamma - 1) U$ is the gas thermal energy, $V$ is the considered plasma volume enclosed by the surface $S$, and the other symbols have their usual meaning. The above expression was derived in \citet{Somov2007} starting from the MHD motion equation and by integrating it over a volume $V$ after a scalar product by $\mathbf{r}$. In the above expression the two terms $\int_V 3p\,dV$ and $\oint_S p\left( \mathbf{r} \cdot \mathbf{dS} \right)$ come from the same volume integral $\int_V \mathbf{r} \cdot \mathbf{\nabla}p\,dV$ which, by using the divergence theorem, has been rewritten as
\begin{eqnarray}
	\int_V \mathbf{r} \cdot \mathbf{\nabla}p\,dV = \int_V \mathbf{\nabla} \cdot \left( p\mathbf{r}\right) dV - \int_V p\, \mathbf{\nabla} \cdot \mathbf{r} \,dV = \\
	= \oint_S p\left( \mathbf{r} \cdot \mathbf{dS} \right) - \int_V 3p \,dV \nonumber
\end{eqnarray}
considering that obviously $\mathbf{\nabla} \cdot \mathbf{r} = 3$. However, for the purposes of the discussion presented here, it is more convenient to rewrite the above theorem as
\begin{eqnarray}
	\int_V \left( \rho v^2-r\frac{dp}{dr}+\,\frac{B^2}{2\mu}\,-\,\frac{G\,M_{\odot}\rho}{r}\right)\,dV = \label{eq: virial1} \\
	= \oint_S \left[ \frac{B^2}{2\mu} \left( \mathbf{r} \cdot \mathbf{dS} \right) - \frac{1}{\mu} \left( \mathbf{B} \cdot \mathbf{r} \right) \left( \mathbf{B} \cdot \mathbf{dS} \right) \right] \nonumber
\end{eqnarray}
by assuming that in the solar corona $\mathbf{r} \cdot \mathbf{\nabla}p \simeq r\, dp/dr$. Now, from the above equation we need to derive a relationship between the magnetic field energy and the gravitational potential energy, thus eliminating the other terms. This can be done as follows: let us assume that $V$ is a spherical shell in the solar corona, hence it is a volume enclosed between two spherical surfaces $S_1$ and $S_2$ concentric with the Sun with radii $r_1$ and $r_2$ (with $r_1 < r_2$), respectively. The left-hand side of Eq.~\ref{eq: virial1} can be approximated by assuming that in the inner corona analyzed in this work both the plasma dynamic pressure $\rho v^2$ and the gas pressure $p = 2n_ek_BT_e$ are negligible (because $\beta \ll 1$) with respect to the magnetic pressure and the gravitational potential (see also Fig.~\ref{Fig1}). Hence, this first term can be approximated as
\begin{equation}
	\int_V \left( \frac{B^2}{2\mu} - \frac{G\,M_\odot \rho}{r}\right) dV \simeq 4 \pi r^2\left( \frac{\langle B^2 \rangle_{4\pi}}{2\mu}-\frac{G\,M_\odot \langle \rho \rangle_{4\pi}}{r} \right) \Delta r \label{eq: virial left term}  
\end{equation}
where $\Delta r = r_2-r_1$, and the symbol $\langle ... \rangle_{4\pi}$ denotes the average obtained at constant altitude $r$ over the whole spherical shell in the corona. On the other hand, on the right-hand side of Eq.~\ref{eq: virial1} over the outer spherical surface $S_2$ it is $\mathbf{r} \cdot \mathbf{dS} = r\,dS$, and $(\mathbf{B} \cdot \mathbf{r})(\mathbf{B} \cdot \mathbf{dS}) = r\,B_r^2 dS$, while over the inner spherical surface $S_1$ it is $\mathbf{r} \cdot \mathbf{dS} = -r\,dS$, and $(\mathbf{B} \cdot \mathbf{r})(\mathbf{B} \cdot \mathbf{dS}) = - r\,B_r^2 dS$. Hence, by splitting the total integral over the surface $S = S_1+S_2$ enclosing the volume $V$ into two integrals, and considering that $B^2 = B_r^2+B_\theta^2+B_\phi^2$, the right-hand side of Eq.~\ref{eq: virial1} becomes equal to
\begin{eqnarray} \nonumber
	-\frac{1}{2\mu} \oint_{S_1} \left[ B^2 (r\,dS) - (2r\,B_r^2 dS) \right] + \\
	+ \frac{1}{2\mu} \oint_{S_2} \left[ B^2 (r\,dS) - (2r\,B_r^2 dS) \right] = \\ \nonumber
	= -\frac{1}{2\mu} \oint_{S_1} r(B_\theta^2 +B_\phi^2-B_r^2) dS + \\ \nonumber
	+ \frac{1}{2\mu} \oint_{S_2} r(B_\theta^2 +B_\phi^2-B_r^2) dS = \label{eq: surface integrals} \\
	= \frac{1}{2\mu} \oint_{S_1} r \left( B_r^2-B_t^2 \right) dS - \frac{1}{2\mu} \oint_{S_2} r \left( B_r^2-B_t^2 \right) dS
\end{eqnarray}
having defined $B_\theta^2 + B_\phi^2 = B_t^2$ as the transversal magnetic field component; the above expressions are similar to those provided by \citet{Wolfson1992} also derived from the virial theorem.

\subsection{Lorentz force}

The expressions in the above integrals correspond to the radial components of the surface Lorentz force. In particular, the Lorentz force per unit volume $F_L$ can be written in general as the divergence of the Maxwell stress tensor \citep[see e.g.][]{Ogilvie2016}:
\begin{equation}
	\mathbf{F_L} = \mathbf{\nabla} \cdot \mathbf{M}\,,\,\,\mathbf{M} = \frac{1}{\mu} \left( \mathbf{BB} -\frac{B^2}{2}\mathbf{I} \right),
\end{equation}
where $\mathbf{I}$ is the identity tensor. By considering a volume $V$ of plasma bounded by a closed surface $S$, the net Lorentz force acting on $V$ can be written by the divergence theorem as
\begin{equation}
	\int_V \mathbf{\nabla} \cdot \mathbf{M}\, dV = \oint_S \mathbf{M} \cdot \mathbf{\hat{n}}\, dS
\end{equation}
where $\mathbf{\hat{n}}$ is the outward unit vector normal to the surface $S$, and $\mathbf{F_S} = \mathbf{M} \cdot \mathbf{\hat{n}}$ is the surface force per unit area \citep[see][]{Spruit2013}. Now, by decomposing in spherical coordinates the magnetic field into radial $B_r$ and transversal $B_t$ components, so that $\mathbf{B} = B_r \mathbf{\hat{r}} + B_t \mathbf{\hat{t}}$ (with $\mathbf{\hat{r}}$ and $\mathbf{\hat{t}}$ the unit radial and transversal vectors), the force $\mathbf{F_S}$ can be written as
\begin{eqnarray} \nonumber
	\mathbf{F_S} = \frac{1}{\mu} \left( \mathbf{B}B_r-\frac{B^2}{2}\mathbf{\hat{r}}\right) = \\ \nonumber
	= \frac{1}{2\mu} \left[ 2\left( B_r \mathbf{\hat{r}} + B_t \mathbf{\hat{t}} \right) B_r  - \left( B_r^2+B_t^2 \right) \mathbf{\hat{r}} \right] = \\ \nonumber
	= \frac{1}{2\mu} \left[ \left( B_r^2-B_t^2\right) \mathbf{\hat{r}} + 2 B_rB_t \mathbf{\hat{t}} \right] = \mathbf{F_{Sr}} + \mathbf{F_{St}}
\end{eqnarray}
with $\mathbf{F_{Sr}} = (B_r^2-B_t^2)/2\mu\,\mathbf{\hat{r}}$ and $\mathbf{F_{St}} = B_r B_t/\mu\,\mathbf{\hat{t}}$ denoting the radial and transversal components of the Lorentz surface force, respectively. Notice that according to the above expression $\mathbf{F_S}$ is directed outward (inward) when $B_r^2 > B_t^2$ ($B_r^2 < B_t^2$), with a positive (negative) value of the radial component due to the predominant magnetic pressure (tension) of the Lorentz force; for $B_t = 0$ the force is only directed upward, while for $B_r =0$ it is only directed downward, as expected. With the above expression for $\mathbf{F_S}$ the surface integrals in Eq.~\ref{eq: surface integrals} can be rewritten more simply as
\begin{equation}
	- \oint_{S_1} \mathbf{F_S} \cdot \mathbf{r}\, dS - \oint_{S_2} \mathbf{F_S} \cdot \mathbf{r}\, dS = - \oint_{S} \mathbf{F_S} \cdot \mathbf{r}\, dS
\end{equation}
having considered that over the inner spherical surface $S_1$ it is $\mathbf{\hat{n}} = -\mathbf{\hat{r}}$, while over the outer surface $S_2$ it is $\mathbf{\hat{n}} = \mathbf{\hat{r}}$.

Hence, the above term corresponding to Eq.~\ref{eq: surface integrals} can be interpreted as the total work done by the Lorentz force applied by the external fields to the volume $V$ of plasma. By averaging over the spherical surfaces $S_1$ and $S_2$ and for $r_1 \sim r_2$ the Eq.~\ref{eq: surface integrals} can be approximated as
\begin{equation}
	\simeq - \frac{1}{2\mu} 4\pi r^2 \langle B_r^2-B_t^2 \rangle_{4\pi} \Delta r. \label{eq: virial right term}
\end{equation}
Finally, by combining Eq.~\ref{eq: virial left term} and Eq.~\ref{eq: virial right term}, and by simplifying the resulting expression, we have
\begin{eqnarray}
	\frac{\langle B^2 \rangle_{4\pi}}{2\mu}-\frac{G\,M_\odot \langle \rho \rangle_{4\pi}}{r} = - \frac{\langle B_r^2-B_t^2 \rangle_{4\pi}}{2\mu}. \label{eq: final virial}
\end{eqnarray}
Considering that $\langle B^2 \rangle_{4\pi} = \langle B_r^2+B_t^2 \rangle_{4\pi} = \langle B_r^2 \rangle_{4\pi} + \langle B_t^2 \rangle_{4\pi}$, and that $\langle B_r^2-B_t^2 \rangle_{4\pi} = \langle B_r^2 \rangle_{4\pi} - \langle B_t^2 \rangle_{4\pi}$, the above equation simplifies into
\begin{equation}
	\frac{\langle B_r^2 \rangle_{4\pi}}{\mu}-\frac{G\,M_\odot \langle \rho \rangle_{4\pi}}{r} = 0 \label{eq: final virial2}
\end{equation}
which allows to derive an expression for the rms value $\langle |B_r| \rangle_{4\pi} = \sqrt{\langle B_r^2 \rangle_{4\pi}}$ of the radial magnetic field which is given by
\begin{equation}
	\langle |B_r| \rangle_{4\pi} = \sqrt{\frac{\mu\, G M_\odot \langle \rho \rangle}{r}} = \frac{1}{2}\,\langle B_{eq} \rangle_{4\pi} \label{eq: absolute radial field}
\end{equation}
according to the definition of $B_{eq}$ given in Eq.~\ref{eq: equipotential field}. 

In conclusion, by averaging over a spherical surface $S$ with radius $r$ concentric with the Sun, one should expect to obtain the value given by the above expression for the rms value at constant altitude of the radial magnetic field component $\langle |B_r| \rangle_{4\pi}$, while in general it should also be $\langle B_r \rangle_{4\pi} = 0$ because the total magnetic flux across $S$ should be zero. By assuming that for the above quantities $\langle ... \rangle_{2\pi} \simeq \langle ... \rangle_{4\pi}$ and denoting hereafter these two averages with the same symbol $\langle ... \rangle$, we have
\begin{equation}
	\langle B_{eq} \rangle = 2 \langle |B_r| \rangle \label{eq: Beq}.
\end{equation}
From Eq.~\ref{eq: equipotential field} this leads to conclude that
\begin{equation}
	B_{pot} = 2 \langle B_{eq} \rangle - B_{eq} = 4 \langle |B_r| \rangle - B_{eq}.
\end{equation}

\subsection{Alfv\'en waves}

The above expression for $B_{eq}$ can be more generalized by considering that in the original formulation of the virial theorem by \citet{Somov2007} (Eq.~\ref{eq: virial1}) there is at least one missing term that can be added for the applications in the solar corona considered here, expressing the additional plasma pressure due to the propagation of \alfven\ waves. In particular, as first provided by \citet{Jacques1977}, the presence of \alfven\ waves corresponds to an additional pressure term $\mathbf{\nabla} \cdot \mathbf{P_w}$, with $\mathbf{P_w}$ denoting the wave stress tensor. This additional pressure is thus anisotropic, and must be considered whether or not the waves are damped. For the specific case of radially propagating \alfven\ waves the general expression for $\mathbf{P_w}$ representing the effect of the waves on the mean flow reduces to \citep{Jacques1977}
\begin{equation}
	\mathbf{P_w} = \frac{1}{2} \varepsilon_w \mathbf{1}
\end{equation}
where $\varepsilon_w = 1/2\,\rho \omega_0^2 a_w^2$ is the wave energy density, and $\omega_0$ and $a_w$ are the wave intrinsic frequency and wave amplitude, respectively. Hence, the wave energy density, also related to the quadratic mean value of magnetic field fluctuations $\delta B^2$ simply by $\varepsilon_w = \delta B^2/(2\mu)$, reduces to an isotropic pressure $1/2\,\varepsilon_w$, and the radial component of the divergence of the wave stress tensor simplifies to
\begin{equation}
	\left( \mathbf{\nabla} \cdot \mathbf{P_w}\right)_r = \frac{1}{2} \frac{d\varepsilon_w}{dr}.
\end{equation}
The above term provides an additional pressure term to be added on the left side of Eq.~\ref{eq: virial1} given by
\begin{equation}
	- \left( \mathbf{\nabla} \cdot \mathbf{P_w}\right)_r \mathbf{r}= -\frac{1}{2}r\frac{d\varepsilon_w}{dr}. \label{eq: pressure term}
\end{equation}
The expression for the radial gradient of the \alfven\ wave energy density $d\varepsilon_w/dr$ (and hence for the magnetic field fluctuations $\delta B^2$) is in general unknown, but a simple relationship was provided for instance by \citet{Holzer1983} who showed that, in the case of nondissipative energy exchange between the \alfven\ waves and the background medium, and in the limit of small Alfv\'en Mach numbers $M_A = v_A/v \ll 1$ usually valid in the inner corona, the magnetic field fluctuations associated with \alfven\ waves are given by
\begin{equation}
	\delta B^2 = \delta B_0^2 \sqrt{\frac{\rho}{\rho_0}}
\end{equation}
with $\delta B_0$ and $\rho_0$ denoting the magnetic field fluctuations and plasma density at the base of the corona ($r = 1$ R$_\odot$). By replacing this expression in the above additional term (Eq.~\ref{eq: pressure term}) we have
\begin{equation}
	- \frac{1}{2}r\,\frac{d\varepsilon_w}{dr} = - \frac{1}{2}r\,\frac{d}{dr} \left(\frac{\delta B^2}{2\mu} \right) = - \frac{1}{4}r\, \frac{\delta B_0^2}{2\mu}\, \frac{1}{\sqrt{\rho_0 \rho}}\,\frac{d\rho}{dr} > 0 
\end{equation}
because in the stationary solar corona it is always $d\rho/dr <0$. A more refined expression for $\delta B^2$ as a function of heliocentric distance has been provided for instance by \citet{Hansteen2012} (see their Eq.~19). What is important here is to notice that the above term to be added on the left side of Eq.~\ref{eq: virial1} is expected to be always positive in the solar corona.

In conclusion, by also considering this term representing the additional pressure due to the \alfven\ waves, the expression previously derived for the rms value $\langle | B_r | \rangle$ (Eq.~\ref{eq: absolute radial field}) can be more generalized as
\begin{equation}
	\langle | B_r | \rangle = \sqrt{ \frac{\mu G M_\odot \langle \rho \rangle}{r} +\mu\, \frac{1}{2}r\,\langle \frac{d\varepsilon_w}{dr} \rangle }.
\end{equation}
Since $d\varepsilon_w/dr$ is negative, the $\langle | B_r | \rangle$ value resulting from the above expression is smaller than the one provided by Eq.~\ref{eq: absolute radial field}, thus increasing the difference between $\langle | B | \rangle$ and $\langle B_{eq} \rangle$ which will be larger in general than a factor of 2.

In summary, the above considerations show that, starting from the virial theorem, by neglecting the dynamic and thermal plasma pressures, and by including the role of the Lorentz force, it is possible to demonstrate that at constant altitude $r$ the average equipartition magnetic field $\langle B_{eq} \rangle$ (defined with Eq.~\ref{eq: equipartition field}) corresponds to half of the rms value of the coronal radial magnetic field $\langle | B_r | \rangle $. Hence, within a factor of 2, a theoretical derivation for $\langle B_{eq} \rangle$ is provided. By also adding the pressure term due to the \alfven\ waves, the resulting expression for $\langle | B_r | \rangle$ is further reduced, thus increasing the gap between the two quantities. On the other hand, it is not possible to provide here a theoretical derivation for the expression of the equipotential magnetic field $B_{pot}$ (Eq.~\ref{eq: equipotential field}); this will hopefully be provided in a future work.

\section{Coronal magnetic field direction} \label{sec: field direction}

Images acquired during a TSE can be also employed to derive information on the direction of the magnetic fields, projected on the POS. In a recent work, \citet{Benjamin2020} demonstrated how images acquired during a TSE can be analyzed to infer the topological properties of coronal magnetic fields by applying a rolling Hough transform to the observations acquired during 14 TSEs covering almost two complete solar cycles. The underlying hypothesis is that in the inner corona where $\beta < 1$ the observed coronal features can be used directly as tracers of the magnetic field inclination. Here, I apply a simpler technique to demonstrate how the inclination of the magnetic fields can be derived in a semi-automated method by analyzing in particular the images that I acquired during the 2017 TSE \citep{Bemporad2020}. The analysis performed here consists of two steps as described below: image filtering, and intensity variance analysis.
\begin{figure*}[t!]
	\centering
	\includegraphics[width=0.9\textwidth]{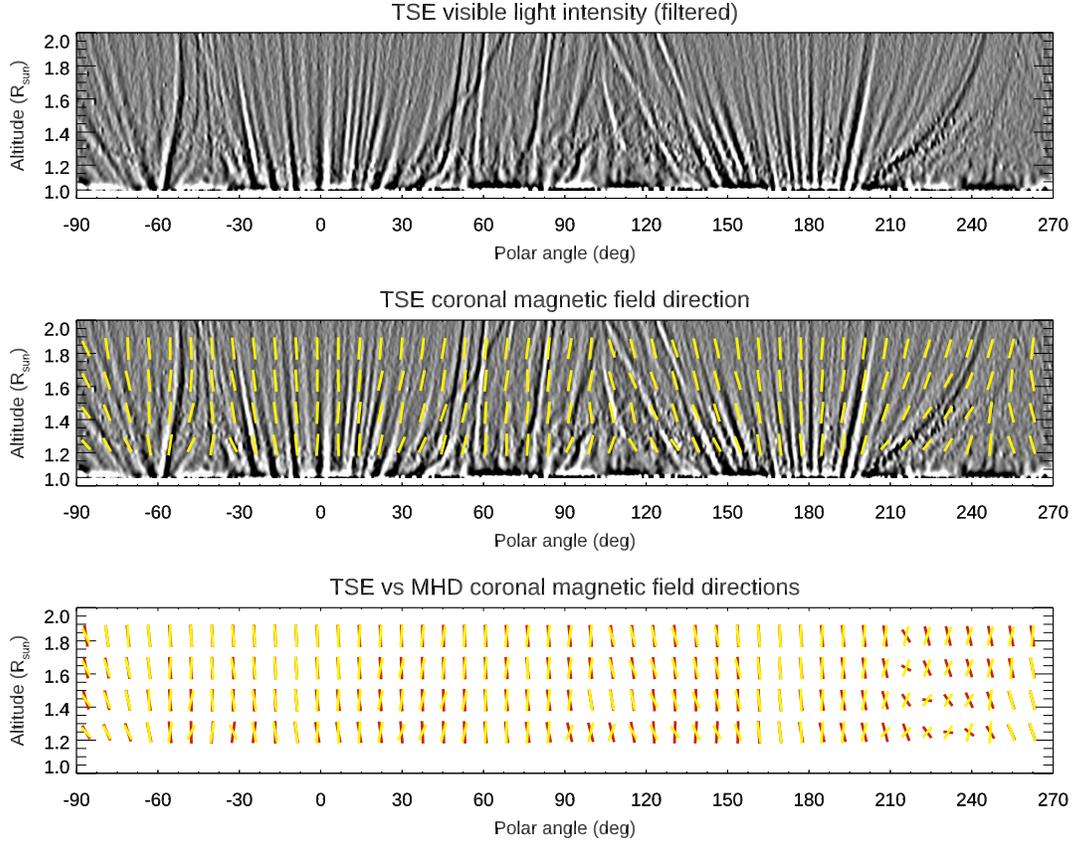}
	\caption{
		Top: polar map of the visible light intensity distribution as obtained from the 2017 TSE observations after filtering to enhance the visibility of fainter coronal features. Middle: direction of the coronal magnetic fields as reconstructed from the observed orientation of the coronal features. Bottom: comparison between the direction of the coronal magnetic fields as reconstructed from TSE images (yellow) and the corresponding direction of coronal magnetic fields as predicted by the MHD numerical simulations (red).
	}
	\label{Fig10}
\end{figure*} 

\subsection{Image filtering}

The first step in the semi-automated analysis performed here consists in the application of image filtering, to enhance the visibility of fainter coronal features in images obtained during the 2017 TSE observations. A significant improvement in the visibility of fainter features \citep[see Fig.~9 in][]{Bemporad2020} is provided already by applying a standard normalizing radial graded filter \citep[NRGF; see][]{Morgan2006}, followed by the application of an EDGE\_DOG filter (a standard bandpass filter based on the subtraction of two copies of the same image obtained after the application of different Gaussian blurrings).

Here I show for the first time that, as an alternative to the NRGF, a significant improvement can be more easily obtained by simply subtracting to each other two copies of the same coronal image rotated each other by an angle $\pm \Delta \theta$ around the center of the Sun. This simple method already allows to "flatten" the coronal image in the radial direction and also to enhance the visibility of fainter coronal features that appear as a "black-and-white" radial pairs. In particular, the image shown in the top panel of Fig.~\ref{Fig10} is obtained with $\Delta \theta = 0.3^\circ$, followed by the application of the EDGE\_DOG filter, and then by conversion from Cartesian to polar coordinates of the filtered image. The conversion is performed here by considering only the coronal region between projected altitudes of 1.05 and 2.0 R$_\odot$, with a radial sampling by 0.00388 R$_\odot$ pixel$^{-1}$ and an angular sampling by 0.111$^\circ$ pixel$^{-1}$.

Notice that the above simple filtering procedure, as well as the one applied in \citet[][]{Bemporad2020}, is based on the analysis of a single image, and not on a combination of coronal images acquired at different times. The same is also true for the NRGF \citep{Morgan2006}, for the improved version of NRGF with Fourier transform \citep{Druckmullerova2011}, and for the more advanced filtering techniques developed for instance by \citet{Druckmuller2006}. All these kinds of "self-filtering" methods, i.e. methods based on the analysis of a single image, cannot introduce artifacts due to the time evolution of the corona. On the other hand, "time-dependent filtering" methods \citep[such as the recently proposed SIRGraF;][]{Patel2022} can introduce artifacts when applied to an observational sequence whose temporal duration is not significantly longer than the evolution time scale of the phenomenon under study, as recently done for instance by \citet{Telloni2022} to claim the observation of a "magnetic switchback" in the solar corona.
\begin{figure*}[!t]
	\centering
	\vspace{-2.8cm}
	\includegraphics[trim={2cm 3cm 2cm 0}, width=0.85\textwidth]{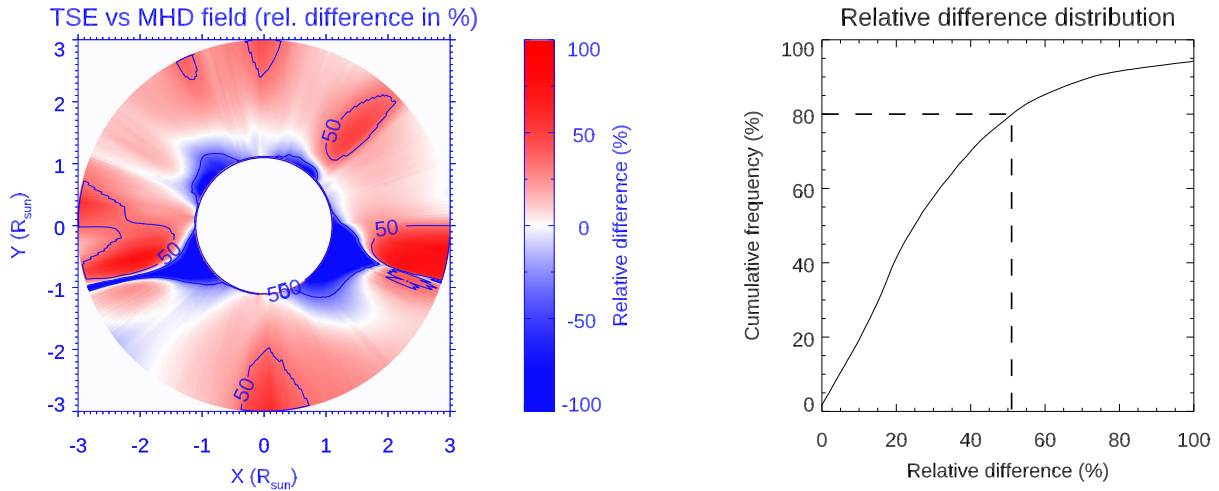}
%	\vspace{-3.2cm}
	\caption{
		Left: POS distribution of the relative differences between the coronal magnetic field strength predicted by the MHD numerical simulations and the equipotential coronal magnetic field strength as derived from TSE measurements. Right: distribution of the relative difference values in the left image.
	}
	\label{Fig11}
\end{figure*} 

\subsection{Intensity variance analysis}

A procedure is developed to identify in a semi-automatic way the inclination angles $\theta_{TSE}$ of coronal features in the filtered image, assumed to be representative of the coronal magnetic field direction projected on the POS. This is done by measuring over a grid of points evenly distributed in the corona the variance of intensity fluctuations at different inclination angles. The main direction of coronal features is then determined point-by-point as the direction corresponding to the minimum intensity variance, building a 2D map of coronal feature orientations on the POS. The resulting $\theta_{TSE}$ map is shown in the middle panel of Fig.~\ref{Fig10}, overplotted on the filtered polar intensity map. The comparison shows the overall agreement between the determined and visible orientations of the coronal features. Moreover, when multiple features with different inclinations appear to be superposed in the intensity polar map, the above method identifies the orientations corresponding to features having the strongest contrast in the filtered image, that are also likely those being located closer to the POS.

We now compare these orientations $\theta_{TSE}$ with the POS inclinations $\theta_{MHD}$ of the coronal magnetic field as provided by the MHD numerical simulations, with $\theta_{MHD} = \arctan{B_\theta/B_r}$ (see also Fig.~\ref{Fig3}). The bottom panel of Fig.~\ref{Fig10} shows the comparison between $\theta_{TSE}$ (yellow lines) and $\theta_{MHD}$ (red lines). In general $\theta_{TSE} \sim \theta_{MHD}$ particularly at higher altitudes ($> 1.5$ R$_\odot$) where the tangential magnetic field component goes to zero and the coronal features (and hence the magnetic fields) become almost radial. On the other hand, in the inner corona ($< 1.5$ R$_\odot$) there are latitudinal regions where greater disagreement is observed, particularly in the South-West quadrant ($180^\circ \leqslant$ PA $\leqslant 270^\circ$). Although this discrepancy may in part be due to LOS integration effects in the optically thin corona, this shows again that, as pointed out with the previous comparison between the observed $n_{TSE}$ and the predicted $n_{MHD}$ coronal densities \citep[and as discussed by][]{Mikic2018}, there are coronal regions that are reconstructed with a lower accuracy by the MHD model.

\section{Summary and conclusions} \label{sec: summary}

In this work I analyze coronal density measurements obtained from a sequence of images acquired during the total solar eclipse (TSE) of August 21st, 2017 \citep{Bemporad2020}. The analysis presented here demonstrates that starting from the assumption of an equipartition principle between the magnetic and gravitational potential energies, it is possible to infer the coronal magnetic field strength based on "standard" plasma density measurements obtained from visible light polarized brightness (pB) observations. Magnetic field measurements derived from this assumption have been compared here with those obtained from one of the most advanced MHD numerical reconstructions currently available, starting from photospheric field measurements \citep{Lionello2009, Mikic2018}, and finding a good agreement. More quantitatively, this comparison is better shown in Fig.~\ref{Fig11}, providing an estimate of the relative difference between the measured and the simulated magnetic fields. This Figure shows that for the 80\% of pixels located in the inner corona ($r < 3$ R$_\odot$) the relative difference between the two magnetic field values is smaller than $\simeq 50\%$. Considering that the magnetic fields reconstructed from the MHD simulation also do not perfectly reproduce in all regions the observed coronal structures \citep[as also discussed by][]{Mikic2018}, these relative differences are very small. In particular, larger differences are present for coronal regions located below 1 \rsun\ at the base of the two South coronal streamers. Also, larger differences are associated with coronal regions located higher up at polar angles multiples of 90$^\circ$ where the pB values (and hence the coronal densities) are measured in this work with larger errors \citep[see also][for more details]{Bemporad2020}, but these differences will be reduced if the same method is applied to a coronal density map not affected by this problem. 

In this regard, it is very important to point out that all the results presented here have been also obtained starting from the pB images kindly provided by \citet{Hanaoka2021} for the same August 21, 2017 TSE, and also for the July 2, 2019 TSE. In both cases, the magnetic fields reconstructed with the coronal densities from the pB images provided by these authors show a very good agreement with the magnetic fields reconstructed with the MHD numerical simulations. These results are not presented here only for brevity. Hence, contrary to the comment by \citet{Parker1984} that "it might appear that nature abhors equipartition of energy between field and fluid", results presented here show that in the inner solar corona some kind of equipartition exists between the magnetic field energy and the gravitational potential energy. In light of the results presented here, part of the efforts carried out by previous authors to measure the coronal magnetic fields with different techniques both from the ground and from space (see Introduction) should be at least reconsidered, because the results of this work provide what could be an alternative and easily accessible method to measure the coronal magnetic fields on the plane of sky.

Moreover, what finally emerges in this work is that the coronal magnetic field should be considered more properly not as a local, but as a global property, because the local magnetic field strength at each latitude depends on the average obtained over all latitudes of the equipotential magnetic field derived from the gravitational potential energy. This conclusion is also partially supported here from the theoretical point of view starting from the MHD scalar virial theorem. In fact, the equipartition and equipotential magnetic fields introduced here are measured starting from the POS distribution of the coronal plasma, but this is only one of the infinite number of possible planes that can be used to observe the corona. Because the derivation of the equipotential magnetic field requires to obtain an average over all the visible latitudes, this property should apply to every possible plane cutting across the solar corona and passing through the solar center. Hence, this leads to conclude that when new magnetic flux emerges in a single region of the photosphere, this modifies not only the local distribution of the plasma and magnetic fields in the surrounding overlying coronal regions (as is usually assumed so far), but also the distribution in the corona as a whole.

It is important to also point out that the results presented here apply only to the stationary solar corona. We now know that the corona is highly dynamic, particularly at the smaller spatial scales being now observed for instance by the instruments on-board Solar Orbiter \citep{Muller2020, Berghmans2021}, but the results presented here show that, considering the whole corona at larger spatial scales and longer time scales, "the coronal plasma naturally seeks out a stable quiescent existence until its environment, by slow changes, forces it to a dynamical transition" \citep{Low1990}. Even though further tests will be needed also based on different numerical modelings, a potentially powerful new tool to routinely measure coronal magnetic fields is presented here. The main uncertainties in this method are certainly related to LOS integration effects (as is usual for remote sensing observations of the solar corona), but the measured densities and magnetic fields are in good agreement with the values predicted by the MHD modeling on the POS, and a good agreement is also found for the POS distributions of other plasma parameters predicted by MHD such as the Alfv\'en speed and the plasma$-\beta$ parameter. Hence, LOS integration effects do not significantly affect the method.
	
In the future it will be very interesting to combine this method with the other already existing measurement techniques briefly described in the Introduction, to optimize and/or cross-calibrate the different measurements. The measurement of coronal magnetic fields in the quasi-stationary corona before and after limb solar eruptions can provide information on the amount of released magnetic energy. Also, given the availability of an impressive number of polarized visible light observations of the inner corona acquired mostly by ground-based coronagraphs \citep[such as the series of Mauna Loa K-Coronagraphs, e.g.][]{Cyr1999, Cyr2015}, this method opens up the possibility of investigating large scale coronal magnetic fields over past solar cycles, deriving new information for instance on the long-term behavior of the solar dynamo \citep{Dikpati2016}. By using one full solar rotation of observations it will be also possible to derive with tomographic reconstructions not only the 3D distribution of coronal densities, but also that of coronal magnetic fields, as done for instance by \citet{Kramar2016}, also reducing possible uncertainties related to the LOS integration effect. Moreover, in the near future the same method could be applied to images that will be acquired during artificial eclipses created by the first formation-flying giant space-based coronagraph ASPIICS on-board the forthcoming ESA PROBA-3 mission \citep{Lamy2010}.

\acknowledgments
The author thanks L. Abbo and C. Benna for their invaluable help and support in the organization of the 2017 TSE observational campaign, making this work possible. The author also thanks Y. Hanaoka for the provision of pB images acquired during the 2017 and 2019 TSEs. The very useful comments, suggestions, and corrections by the anonymous Referee are also acknowledged.

\bibliographystyle{aasjournal}
\bibliography{biblio}{}

\begin{thebibliography}{}
\expandafter\ifx\csname natexlab\endcsname\relax\def\natexlab#1{#1}\fi
\providecommand{\url}[1]{\href{#1}{#1}}
\providecommand{\dodoi}[1]{doi:~\href{http://doi.org/#1}{\nolinkurl{#1}}}
\providecommand{\doeprint}[1]{\href{http://ascl.net/#1}{\nolinkurl{http://ascl.net/#1}}}
\providecommand{\doarXiv}[1]{\href{https://arxiv.org/abs/#1}{\nolinkurl{https://arxiv.org/abs/#1}}}

\bibitem[{{Altschuler} \& {Newkirk}(1969)}]{Altschuler1969}
{Altschuler}, M.~D., \& {Newkirk}, G. 1969, \solphys, 9, 131,
  \dodoi{10.1007/BF00145734}

\bibitem[{{Ballester} \& {Kleczek}(1984)}]{Ballester1984}
{Ballester}, J.~L., \& {Kleczek}, J. 1984, \solphys, 90, 37,
  \dodoi{10.1007/BF00153783}

\bibitem[{{Beckers}(1971)}]{Beckers1971}
{Beckers}, J.~M. 1971, in Solar Magnetic Fields, ed. R.~{Howard}, Vol.~43, 3

\bibitem[{{Bemporad}(2008)}]{Bemporad2008}
{Bemporad}, A. 2008, \apj, 689, 572, \dodoi{10.1086/592377}

\bibitem[{{Bemporad}(2017)}]{Bemporad2017}
---. 2017, \apj, 846, 86, \dodoi{10.3847/1538-4357/aa7de4}

\bibitem[{{Bemporad}(2020)}]{Bemporad2020}
---. 2020, \apj, 904, 178, \dodoi{10.3847/1538-4357/abc482}

\bibitem[{{Bemporad} {et~al.}(2021){Bemporad}, {Giordano}, {Zangrilli}, \&
  {Frassati}}]{Bemporad2021}
{Bemporad}, A., {Giordano}, S., {Zangrilli}, L., \& {Frassati}, F. 2021, \aap,
  654, A58, \dodoi{10.1051/0004-6361/202141276}

\bibitem[{{Berghmans} {et~al.}(2021){Berghmans}, {Auch{\`e}re}, {Long},
  {Soubri{\'e}}, {Mierla}, {Zhukov}, {Sch{\"u}hle}, {Antolin}, {Harra},
  {Parenti}, {Podladchikova}, {Aznar Cuadrado}, {Buchlin}, {Dolla}, {Verbeeck},
  {Gissot}, {Teriaca}, {Haberreiter}, {Katsiyannis}, {Rodriguez}, {Kraaikamp},
  {Smith}, {Stegen}, {Rochus}, {Halain}, {Jacques}, {Thompson}, \&
  {Inhester}}]{Berghmans2021}
{Berghmans}, D., {Auch{\`e}re}, F., {Long}, D.~M., {et~al.} 2021, \aap, 656,
  L4, \dodoi{10.1051/0004-6361/202140380}

\bibitem[{{Boe} {et~al.}(2020){Boe}, {Habbal}, \&
  {Druckm{\"u}ller}}]{Benjamin2020}
{Boe}, B., {Habbal}, S., \& {Druckm{\"u}ller}, M. 2020, \apj, 895, 123,
  \dodoi{10.3847/1538-4357/ab8ae6}

\bibitem[{{Cranmer} {et~al.}(1999){Cranmer}, {Field}, \& {Kohl}}]{Cranmer1999}
{Cranmer}, S.~R., {Field}, G.~B., \& {Kohl}, J.~L. 1999, \apj, 518, 937,
  \dodoi{10.1086/307330}

\bibitem[{{Cranmer} \& {van Ballegooijen}(2005)}]{Cranmer2005}
{Cranmer}, S.~R., \& {van Ballegooijen}, A.~A. 2005, \apjs, 156, 265,
  \dodoi{10.1086/426507}

\bibitem[{{Dikpati} {et~al.}(2016){Dikpati}, {Suresh}, \&
  {Burkepile}}]{Dikpati2016}
{Dikpati}, M., {Suresh}, A., \& {Burkepile}, J. 2016, \solphys, 291, 339,
  \dodoi{10.1007/s11207-015-0831-8}

\bibitem[{{Dolei} {et~al.}(2014){Dolei}, {Bemporad}, \& {Spadaro}}]{Dolei2014}
{Dolei}, S., {Bemporad}, A., \& {Spadaro}, D. 2014, \aap, 562, A74,
  \dodoi{10.1051/0004-6361/201321041}

\bibitem[{{Druckm{\"u}ller} {et~al.}(2006){Druckm{\"u}ller}, {Ru{\v{s}}in}, \&
  {Minarovjech}}]{Druckmuller2006}
{Druckm{\"u}ller}, M., {Ru{\v{s}}in}, V., \& {Minarovjech}, M. 2006,
  Contributions of the Astronomical Observatory Skalnate Pleso, 36, 131

\bibitem[{{Druckm{\"u}llerov{\'a}} {et~al.}(2011){Druckm{\"u}llerov{\'a}},
  {Morgan}, \& {Habbal}}]{Druckmullerova2011}
{Druckm{\"u}llerov{\'a}}, H., {Morgan}, H., \& {Habbal}, S.~R. 2011, \apj, 737,
  88, \dodoi{10.1088/0004-637X/737/2/88}

\bibitem[{{Dulk} \& {McLean}(1978)}]{Dulk1978}
{Dulk}, G.~A., \& {McLean}, D.~J. 1978, \solphys, 57, 279,
  \dodoi{10.1007/BF00160102}

\bibitem[{{Feng} {et~al.}(2012){Feng}, {Yang}, {Xiang}, {Jiang}, {Ma}, {Wu},
  {Zhong}, \& {Zhou}}]{Feng2012}
{Feng}, X., {Yang}, L., {Xiang}, C., {et~al.} 2012, \solphys, 279, 207,
  \dodoi{10.1007/s11207-012-9969-9}

\bibitem[{{Flyer} {et~al.}(2005){Flyer}, {Fornberg}, {Thomas}, \&
  {Low}}]{Flyer2005}
{Flyer}, N., {Fornberg}, B., {Thomas}, S., \& {Low}, B.~C. 2005, \apj, 631,
  1239, \dodoi{10.1086/432661}

\bibitem[{{Gibson} {et~al.}(2016){Gibson}, {Kucera}, {White}, {Dove}, {Fan},
  {Forland}, {Rachmeler}, {Downs}, \& {Reeves}}]{Gibson2016}
{Gibson}, S., {Kucera}, T., {White}, S., {et~al.} 2016, Frontiers in Astronomy
  and Space Sciences, 3, 8, \dodoi{10.3389/fspas.2016.00008}

\bibitem[{{Gibson} {et~al.}(1999){Gibson}, {Fludra}, {Bagenal}, {Biesecker},
  {del Zanna}, \& {Bromage}}]{Gibson1999}
{Gibson}, S.~E., {Fludra}, A., {Bagenal}, F., {et~al.} 1999, \jgr, 104, 9691,
  \dodoi{10.1029/98JA02681}

\bibitem[{{Gibson} {et~al.}(2017){Gibson}, {Rachmeler}, \&
  {White}}]{Gibson2017}
{Gibson}, S.~E., {Rachmeler}, L.~A., \& {White}, S.~M. 2017, Frontiers in
  Astronomy and Space Sciences, 4, 3, \dodoi{10.3389/fspas.2017.00003}

\bibitem[{{Gopalswamy} \& {Yashiro}(2011)}]{Gopalswamy2011}
{Gopalswamy}, N., \& {Yashiro}, S. 2011, \apjl, 736, L17,
  \dodoi{10.1088/2041-8205/736/1/L17}

\bibitem[{{Haerendel} \& {Berger}(2011)}]{Haerendel2011}
{Haerendel}, G., \& {Berger}, T. 2011, \apj, 731, 82,
  \dodoi{10.1088/0004-637X/731/2/82}

\bibitem[{{Hanaoka} {et~al.}(2021){Hanaoka}, {Sakai}, \&
  {Takahashi}}]{Hanaoka2021}
{Hanaoka}, Y., {Sakai}, Y., \& {Takahashi}, K. 2021, \solphys, 296, 158,
  \dodoi{10.1007/s11207-021-01907-0}

\bibitem[{{Hansteen} \& {Velli}(2012)}]{Hansteen2012}
{Hansteen}, V.~H., \& {Velli}, M. 2012, \ssr, 172, 89,
  \dodoi{10.1007/s11214-012-9887-z}

\bibitem[{{Holzer} {et~al.}(1983){Holzer}, {Fla}, \& {Leer}}]{Holzer1983}
{Holzer}, T.~E., {Fla}, T., \& {Leer}, E. 1983, \apj, 275, 808,
  \dodoi{10.1086/161576}

\bibitem[{{Hundhausen} {et~al.}(1981){Hundhausen}, {Hundhausen}, \&
  {Zweibel}}]{Hundhausen1981}
{Hundhausen}, J.~R., {Hundhausen}, A.~J., \& {Zweibel}, E.~G. 1981, \jgr, 86,
  11,117, \dodoi{10.1029/JA086iA13p11117}

\bibitem[{{Hundhausen} \& {Low}(1994)}]{Hundhausen1994}
{Hundhausen}, J.~R., \& {Low}, B.~C. 1994, \apj, 429, 876,
  \dodoi{10.1086/174372}

\bibitem[{{Jacques}(1977)}]{Jacques1977}
{Jacques}, S.~A. 1977, \apj, 215, 942, \dodoi{10.1086/155430}

\bibitem[{{Kramar} {et~al.}(2016){Kramar}, {Lin}, \& {Tomczyk}}]{Kramar2016}
{Kramar}, M., {Lin}, H., \& {Tomczyk}, S. 2016, \apjl, 819, L36,
  \dodoi{10.3847/2041-8205/819/2/L36}

\bibitem[{{Kuridze} {et~al.}(2019){Kuridze}, {Mathioudakis}, {Morgan},
  {Oliver}, {Kleint}, {Zaqarashvili}, {Reid}, {Koza}, {L{\"o}fdahl},
  {Hillberg}, {Kukhianidze}, \& {Hanslmeier}}]{Kuridze2019}
{Kuridze}, D., {Mathioudakis}, M., {Morgan}, H., {et~al.} 2019, \apj, 874, 126,
  \dodoi{10.3847/1538-4357/ab08e9}

\bibitem[{{Laming} \& {Feldman}(2003)}]{Laming2003}
{Laming}, J.~M., \& {Feldman}, U. 2003, \apj, 591, 1257, \dodoi{10.1086/375395}

\bibitem[{{Lamy} {et~al.}(2010){Lamy}, {Dam{\'e}}, {Viv{\`e}s}, \&
  {Zhukov}}]{Lamy2010}
{Lamy}, P., {Dam{\'e}}, L., {Viv{\`e}s}, S., \& {Zhukov}, A. 2010, in Society
  of Photo-Optical Instrumentation Engineers (SPIE) Conference Series, Vol.
  7731, Space Telescopes and Instrumentation 2010: Optical, Infrared, and
  Millimeter Wave, ed. J.~{Oschmann}, Jacobus~M., M.~C. {Clampin}, \& H.~A.
  {MacEwen}, 773118, \dodoi{10.1117/12.858247}

\bibitem[{{Lamy} {et~al.}(2017){Lamy}, {Viv{\`e}s}, {Curdt}, {Dam{\'e}},
  {Davila}, {Defise}, {Fineschi}, {Heinzel}, {Howard}, {Kuzin}, {Schmutz},
  {Tsinganos}, \& {Zhukov}}]{Lamy2017}
{Lamy}, P.~L., {Viv{\`e}s}, S., {Curdt}, W., {et~al.} 2017, in Society of
  Photo-Optical Instrumentation Engineers (SPIE) Conference Series, Vol. 10565,
  Society of Photo-Optical Instrumentation Engineers (SPIE) Conference Series,
  105650T, \dodoi{10.1117/12.2309188}

\bibitem[{{Leblanc} {et~al.}(1998){Leblanc}, {Dulk}, \&
  {Bougeret}}]{Leblanc1998}
{Leblanc}, Y., {Dulk}, G.~A., \& {Bougeret}, J.-L. 1998, \solphys, 183, 165,
  \dodoi{10.1023/A:1005049730506}

\bibitem[{{Lemaire}(2011)}]{Lemaire2011}
{Lemaire}, J.~F. 2011, arXiv e-prints, arXiv:1112.3850.
\newblock \doarXiv{1112.3850}

\bibitem[{{Lemaire} \& {Stegen}(2016)}]{Lemaire2016}
{Lemaire}, J.~F., \& {Stegen}, K. 2016, \solphys, 291, 3659,
  \dodoi{10.1007/s11207-016-1001-3}

\bibitem[{{Lin} {et~al.}(2004){Lin}, {Kuhn}, \& {Coulter}}]{Lin2004}
{Lin}, H., {Kuhn}, J.~R., \& {Coulter}, R. 2004, \apjl, 613, L177,
  \dodoi{10.1086/425217}

\bibitem[{{Lin} {et~al.}(2000){Lin}, {Penn}, \& {Tomczyk}}]{Lin2000}
{Lin}, H., {Penn}, M.~J., \& {Tomczyk}, S. 2000, \apjl, 541, L83,
  \dodoi{10.1086/312900}

\bibitem[{{Linker} {et~al.}(1999){Linker}, {Miki{\'c}}, {Biesecker}, {Forsyth},
  {Gibson}, {Lazarus}, {Lecinski}, {Riley}, {Szabo}, \&
  {Thompson}}]{Linker1999}
{Linker}, J.~A., {Miki{\'c}}, Z., {Biesecker}, D.~A., {et~al.} 1999, \jgr, 104,
  9809, \dodoi{10.1029/1998JA900159}

\bibitem[{{Lionello} {et~al.}(2009){Lionello}, {Linker}, \&
  {Miki{\'c}}}]{Lionello2009}
{Lionello}, R., {Linker}, J.~A., \& {Miki{\'c}}, Z. 2009, \apj, 690, 902,
  \dodoi{10.1088/0004-637X/690/1/902}

\bibitem[{{Low}(1975)}]{Low1975}
{Low}, B.~C. 1975, \apj, 197, 251, \dodoi{10.1086/153508}

\bibitem[{{Low}(1982)}]{Low1982}
---. 1982, \apj, 263, 952, \dodoi{10.1086/160563}

\bibitem[{{Low}(1990)}]{Low1990}
---. 1990, \araa, 28, 491, \dodoi{10.1146/annurev.aa.28.090190.002423}

\bibitem[{{Low} \& {Smith}(1993)}]{Low1993}
{Low}, B.~C., \& {Smith}, D.~F. 1993, \apj, 410, 412, \dodoi{10.1086/172758}

\bibitem[{{Mackay} \& {Yeates}(2012)}]{Mackay2012}
{Mackay}, D.~H., \& {Yeates}, A.~R. 2012, Living Reviews in Solar Physics, 9,
  6, \dodoi{10.12942/lrsp-2012-6}

\bibitem[{{Mancuso} \& {Garzelli}(2013)}]{Mancuso2013}
{Mancuso}, S., \& {Garzelli}, M.~V. 2013, \aap, 553, A100,
  \dodoi{10.1051/0004-6361/201220319}

\bibitem[{{McComas} {et~al.}(1996){McComas}, {Hoogeveen}, {Gosling},
  {Phillips}, {Neugebauer}, {Balogh}, \& {Forsyth}}]{McComas1996}
{McComas}, D.~J., {Hoogeveen}, G.~W., {Gosling}, J.~T., {et~al.} 1996, \aap,
  316, 368

\bibitem[{{Miki{\'c}} {et~al.}(2018){Miki{\'c}}, {}, {Downs}, {Linker},
  {Caplan}, {Mackay}, {Upton}, {Riley}, {Lionello}, {T{\"o}r{\"o}k}, {Titov},
  {Wijaya}, {Druckm{\"u}ller}, {Pasachoff}, \& {Carlos}}]{Mikic2018}
{Miki{\'c}}, {}, Z., {Downs}, C., {et~al.} 2018, Nature Astronomy, 2, 913,
  \dodoi{10.1038/s41550-018-0562-5}

\bibitem[{{Miki{\'c}} {et~al.}(2007){Miki{\'c}}, {Linker}, {Lionello}, {Riley},
  \& {Titov}}]{Mikic2007}
{Miki{\'c}}, Z., {Linker}, J.~A., {Lionello}, R., {Riley}, P., \& {Titov}, V.
  2007, in Astronomical Society of the Pacific Conference Series, Vol. 370,
  Solar and Stellar Physics Through Eclipses, ed. O.~{Demircan}, S.~O. {Selam},
  \& B.~{Albayrak}, 299

\bibitem[{{Miki{\'c}} {et~al.}(1999){Miki{\'c}}, {Linker}, {Schnack},
  {Lionello}, \& {Tarditi}}]{Mikic1999}
{Miki{\'c}}, Z., {Linker}, J.~A., {Schnack}, D.~D., {Lionello}, R., \&
  {Tarditi}, A. 1999, Physics of Plasmas, 6, 2217, \dodoi{10.1063/1.873474}

\bibitem[{{Morgan} {et~al.}(2006){Morgan}, {Habbal}, \& {Woo}}]{Morgan2006}
{Morgan}, H., {Habbal}, S.~R., \& {Woo}, R. 2006, \solphys, 236, 263,
  \dodoi{10.1007/s11207-006-0113-6}

\bibitem[{{M{\"u}ller} {et~al.}(2020){M{\"u}ller}, {St. Cyr}, {Zouganelis},
  {Gilbert}, {Marsden}, {Nieves-Chinchilla}, {Antonucci}, {Auch{\`e}re},
  {Berghmans}, {Horbury}, {Howard}, {Krucker}, {Maksimovic}, {Owen}, {Rochus},
  {Rodriguez-Pacheco}, {Romoli}, {Solanki}, {Bruno}, {Carlsson}, {Fludra},
  {Harra}, {Hassler}, {Livi}, {Louarn}, {Peter}, {Sch{\"u}hle}, {Teriaca}, {del
  Toro Iniesta}, {Wimmer-Schweingruber}, {Marsch}, {Velli}, {De Groof},
  {Walsh}, \& {Williams}}]{Muller2020}
{M{\"u}ller}, D., {St. Cyr}, O.~C., {Zouganelis}, I., {et~al.} 2020, \aap, 642,
  A1, \dodoi{10.1051/0004-6361/202038467}

\bibitem[{{Nakagawa} \& {Raadu}(1972)}]{Nakagawa1972}
{Nakagawa}, Y., \& {Raadu}, M.~A. 1972, \solphys, 25, 127,
  \dodoi{10.1007/BF00155751}

\bibitem[{{Ogilvie}(2016)}]{Ogilvie2016}
{Ogilvie}, G.~I. 2016, Journal of Plasma Physics, 82, 205820301,
  \dodoi{10.1017/S0022377816000489}

\bibitem[{{Paraschiv} {et~al.}(2015){Paraschiv}, {Bemporad}, \&
  {Sterling}}]{Paraschiv2015}
{Paraschiv}, A.~R., {Bemporad}, A., \& {Sterling}, A.~C. 2015, \aap, 579, A96,
  \dodoi{10.1051/0004-6361/201525671}

\bibitem[{{Parker}(1984)}]{Parker1984}
{Parker}, E.~N. 1984, \apj, 283, 343, \dodoi{10.1086/162312}

\bibitem[{{Patel} {et~al.}(2022){Patel}, {Majumdar}, {Pant}, \&
  {Banerjee}}]{Patel2022}
{Patel}, R., {Majumdar}, S., {Pant}, V., \& {Banerjee}, D. 2022, \solphys, 297,
  27, \dodoi{10.1007/s11207-022-01957-y}

\bibitem[{{Patzold} {et~al.}(1987){Patzold}, {Bird}, {Volland}, {Levy},
  {Seidel}, \& {Stelzried}}]{Patzold1987}
{Patzold}, M., {Bird}, M.~K., {Volland}, H., {et~al.} 1987, \solphys, 109, 91,
  \dodoi{10.1007/BF00167401}

\bibitem[{{Rachmeler} {et~al.}(2013){Rachmeler}, {Gibson}, {Dove}, {DeVore}, \&
  {Fan}}]{Rachmeler2013}
{Rachmeler}, L.~A., {Gibson}, S.~E., {Dove}, J.~B., {DeVore}, C.~R., \& {Fan},
  Y. 2013, \solphys, 288, 617, \dodoi{10.1007/s11207-013-0325-5}

\bibitem[{{Raymond} {et~al.}(1998){Raymond}, {Suleiman}, {Kohl}, \&
  {Noci}}]{Raymond1998b}
{Raymond}, J.~C., {Suleiman}, R., {Kohl}, J.~L., \& {Noci}, G. 1998, \ssr, 85,
  283, \dodoi{10.1023/A:1005162803316}

\bibitem[{{Reisenfeld} {et~al.}(1999){Reisenfeld}, {McComas}, \&
  {Steinberg}}]{Reisenfeld1999}
{Reisenfeld}, D.~B., {McComas}, D.~J., \& {Steinberg}, J.~T. 1999, \grl, 26,
  1805, \dodoi{10.1029/1999GL900368}

\bibitem[{{Richardson} {et~al.}(2003){Richardson}, {Wang}, \&
  {Burlaga}}]{Richardson2003}
{Richardson}, J.~D., {Wang}, C., \& {Burlaga}, L.~F. 2003, \grl, 30, 2207,
  \dodoi{10.1029/2003GL018253}

\bibitem[{{Riley} {et~al.}(2006){Riley}, {Linker}, {Miki{\'c}}, {Lionello},
  {Ledvina}, \& {Luhmann}}]{Riley2006}
{Riley}, P., {Linker}, J.~A., {Miki{\'c}}, Z., {et~al.} 2006, \apj, 653, 1510,
  \dodoi{10.1086/508565}

\bibitem[{{Saito} {et~al.}(1977){Saito}, {Poland}, \& {Munro}}]{Saito1977}
{Saito}, K., {Poland}, A.~I., \& {Munro}, R.~H. 1977, \solphys, 55, 121,
  \dodoi{10.1007/BF00150879}

\bibitem[{{Somov}(2007)}]{Somov2007}
{Somov}, B.~V. 2007, {Plasma Astrophysics, Part II: Reconnection and Flares}

\bibitem[{{Spruit}(2013)}]{Spruit2013}
{Spruit}, H.~C. 2013, arXiv e-prints, arXiv:1301.5572.
\newblock \doarXiv{1301.5572}

\bibitem[{{St. Cyr} {et~al.}(1999){St. Cyr}, {Burkepile}, {Hundhausen}, \&
  {Lecinski}}]{Cyr1999}
{St. Cyr}, O.~C., {Burkepile}, J.~T., {Hundhausen}, A.~J., \& {Lecinski}, A.~R.
  1999, \jgr, 104, 12493, \dodoi{10.1029/1999JA900045}

\bibitem[{{St. Cyr} {et~al.}(2015){St. Cyr}, {Flint}, {Xie}, {Webb},
  {Burkepile}, {Lecinski}, {Quirk}, \& {Stanger}}]{Cyr2015}
{St. Cyr}, O.~C., {Flint}, Q.~A., {Xie}, H., {et~al.} 2015, \solphys, 290,
  2951, \dodoi{10.1007/s11207-015-0780-2}

\bibitem[{{Stenflo}(1978)}]{Stenflo1978}
{Stenflo}, J.~O. 1978, Reports on Progress in Physics, 41, 865,
  \dodoi{10.1088/0034-4885/41/6/002}

\bibitem[{{Sun} \& {Hu}(2005)}]{Sun2005}
{Sun}, S.~J., \& {Hu}, Y.~Q. 2005, Journal of Geophysical Research (Space
  Physics), 110, A05102, \dodoi{10.1029/2004JA010905}

\bibitem[{{Susino} {et~al.}(2015){Susino}, {Bemporad}, \&
  {Mancuso}}]{Susino2015}
{Susino}, R., {Bemporad}, A., \& {Mancuso}, S. 2015, \apj, 812, 119,
  \dodoi{10.1088/0004-637X/812/2/119}

\bibitem[{{Telloni} {et~al.}(2022){Telloni}, {Zank}, {Stangalini}, {Downs},
  {Liang}, {Nakanotani}, {Andretta}, {Antonucci}, {Sorriso-Valvo}, {Adhikari},
  {Zhao}, {Marino}, {Susino}, {Grimani}, {Fabi}, {D'Amicis}, {Perrone},
  {Bruno}, {Carbone}, {Mancuso}, {Romoli}, {Deppo}, {Fineschi}, {Heinzel},
  {Moses}, {Naletto}, {Nicolini}, {Spadaro}, {Teriaca}, {Frassati}, {Jerse},
  {Landini}, {Pancrazzi}, {Russano}, {Sasso}, {Biondo}, {Burtovoi}, {Capuano},
  {Casini}, {Casti}, {Chioetto}, {Leo}, {Giarrusso}, {Liberatore}, {Berghmans},
  {Auch{\`e}re}, {Cuadrado}, {Chitta}, {Harra}, {Kraaikamp}, {Long}, {Mandal},
  {Parenti}, {Pelouze}, {Peter}, {Rodriguez}, {Sch{\"u}hle}, {Schwanitz},
  {Smith}, {Verbeeck}, \& {Zhukov}}]{Telloni2022}
{Telloni}, D., {Zank}, G.~P., {Stangalini}, M., {et~al.} 2022, \apjl, 936, L25,
  \dodoi{10.3847/2041-8213/ac8104}

\bibitem[{{Tomczyk} {et~al.}(2008){Tomczyk}, {Card}, {Darnell}, {Elmore},
  {Lull}, {Nelson}, {Streander}, {Burkepile}, {Casini}, \&
  {Judge}}]{Tomczyk2008}
{Tomczyk}, S., {Card}, G.~L., {Darnell}, T., {et~al.} 2008, \solphys, 247, 411,
  \dodoi{10.1007/s11207-007-9103-6}

\bibitem[{{Tomczyk} {et~al.}(2022){Tomczyk}, {Burkepile}, {De Wijn}, {Gibson},
  {Gilbert}, {Kolinski}, {Landi}, {Lin}, {DeLuca}, {Martinez Pillet}, \&
  {Zhang}}]{Tomczyk2022}
{Tomczyk}, S., {Burkepile}, J., {De Wijn}, A., {et~al.} 2022, in The Third
  Triennial Earth-Sun Summit (TESS, Vol.~54, 2022n7i121p01

\bibitem[{{Trujillo Bueno} \& {del Pino Alem{\'a}n}(2022)}]{Trujillo2022}
{Trujillo Bueno}, J., \& {del Pino Alem{\'a}n}, T. 2022, \araa, 60, 415,
  \dodoi{10.1146/annurev-astro-041122-031043}

\bibitem[{{van de Hulst}(1950)}]{Hulst1950}
{van de Hulst}, H.~C. 1950, \bain, 11, 135

\bibitem[{{Wang} \& {Davila}(2014)}]{Wang2014}
{Wang}, T., \& {Davila}, J.~M. 2014, \solphys, 289, 3723,
  \dodoi{10.1007/s11207-014-0556-0}

\bibitem[{{Wang} \& {Sheeley}(1990)}]{Wang1990}
{Wang}, Y.~M., \& {Sheeley}, N.~R., J. 1990, \apj, 355, 726,
  \dodoi{10.1086/168805}

\bibitem[{{Wiegelmann} {et~al.}(2005){Wiegelmann}, {Lagg}, {Solanki},
  {Inhester}, \& {Woch}}]{Wiegelmann2005}
{Wiegelmann}, T., {Lagg}, A., {Solanki}, S.~K., {Inhester}, B., \& {Woch}, J.
  2005, \aap, 433, 701, \dodoi{10.1051/0004-6361:20042421}

\bibitem[{{Wiegelmann} {et~al.}(2017){Wiegelmann}, {Petrie}, \&
  {Riley}}]{Wiegelmann2017}
{Wiegelmann}, T., {Petrie}, G. J.~D., \& {Riley}, P. 2017, \ssr, 210, 249,
  \dodoi{10.1007/s11214-015-0178-3}

\bibitem[{{Wolfson} \& {Low}(1992)}]{Wolfson1992}
{Wolfson}, R., \& {Low}, B.~C. 1992, \apj, 391, 353, \dodoi{10.1086/171350}

\bibitem[{{Wu} {et~al.}(1990){Wu}, {Sun}, {Chang}, {Hagyard}, \&
  {Gary}}]{Wu1990}
{Wu}, S.~T., {Sun}, M.~T., {Chang}, H.~M., {Hagyard}, M.~J., \& {Gary}, G.~A.
  1990, \apj, 362, 698, \dodoi{10.1086/169307}

\bibitem[{{Yang} {et~al.}(2020){Yang}, {Bethge}, {Tian}, {Tomczyk}, {Morton},
  {Del Zanna}, {McIntosh}, {Karak}, {Gibson}, {Samanta}, {He}, {Chen}, \&
  {Wang}}]{Yang2020}
{Yang}, Z., {Bethge}, C., {Tian}, H., {et~al.} 2020, Science, 369, 694,
  \dodoi{10.1126/science.abb4462}

\bibitem[{{Zapi{\'o}r} \& {Mart{\'\i}nez-G{\'o}mez}(2016)}]{Zapior2016}
{Zapi{\'o}r}, M., \& {Mart{\'\i}nez-G{\'o}mez}, D. 2016, \apj, 817, 123,
  \dodoi{10.3847/0004-637X/817/2/123}

\bibitem[{{Zapi{\'o}r} \& {Rudawy}(2012)}]{Zapior2012}
{Zapi{\'o}r}, M., \& {Rudawy}, P. 2012, \solphys, 280, 445,
  \dodoi{10.1007/s11207-012-0072-z}

\end{thebibliography}

\end{document}